\documentclass[fleqn]{revtex4}

\usepackage{amsmath}
\usepackage{amssymb}
\usepackage{graphicx}
\usepackage{afterpage}

\DeclareMathOperator{\tr}{Tr}

\def\slash#1{\setbox0=\hbox{$#1$}               
        \dimen0=\wd0                            
        \setbox1=\hbox{/} \dimen1=\wd1          
        \ifdim\dimen0>\dimen1                   
        \rlap{\hbox to \dimen0{\hfil/\hfil}}    
        #1                                      
        \else                                   
        \rlap{\hbox to \dimen1{\hfil$#1$\hfil}} 
        /                                       
        \fi}                                    %

\begin{document}

\title{Gluonic Pole Cross Sections and Single Spin Asymmetries\\ 
in Hadron-Hadron Scattering}

\author{C.J. Bomhof}
\email{cbomhof@nat.vu.nl}
\affiliation{
Department of Physics and Astronomy, Vrije Universiteit Amsterdam,\\
NL-1081 HV Amsterdam, the Netherlands}

\author{P.J. Mulders}
\email{mulders@few.vu.nl}
\affiliation{
Department of Physics and Astronomy, Vrije Universiteit Amsterdam,\\
NL-1081 HV Amsterdam, the Netherlands}

\begin{abstract}
The gauge-links connecting the parton field operators in the hadronic 
matrix elements appearing in the transverse momentum dependent
distribution functions give rise to $T$-odd effects.
Due to the process-dependence of the gauge-links the $T$-odd distribution 
functions appear with different pre\-factors.
A consequence is that in the description of single spin asymmetries the parton distribution and fragmentation functions are convoluted with gluonic pole cross sections rather than the basic partonic cross sections.
In this paper we calculate the gluonic pole cross sections encountered in single spin asymmetries in hadron-hadron scattering.
The case of back-to-back pion production in polarized proton-proton scattering is worked out explicitly. 
It is shown how $T$-odd gluon distribution functions originating from gluonic pole matrix elements appear in twofold.
\end{abstract}
\date{\today}
\maketitle

\section{Introduction}

In recent years there has been increasing interest in
the theoretical description of single spin asymmetries (SSA),
which in some processes can grow to several tens of percents in certain 
kinematical regimes~\cite{Adams:1991rw,Adams:1991cs,Bravar:1996ki,
Airapetian:2001eg,Adler:2003pb,Adams:2003fx,Airapetian:2004tw}.
Single spin asymmetries have a characteristic behavior under time
reversal, being time-reversal odd ($T$-odd). 
Interest has increased chiefly because SSA may provide the simplest access to the transverse spin distribution function of quarks in nucleons, 
which cannot be probed 
in inclusive (polarized) deep inelastic scattering. 
The chiral-odd nature of the transverse spin distribution
requires a scattering process with at least two observed hadrons. 
For spin 0 and spin $\frac{1}{2}$ hadrons, the $T$-odd effects require
measurement of azimuthal asymmetries in which the quark transverse
momentum plays a role. The $T$-odd effects vanish upon 
integration over the transverse momentum, 
but not in transverse momentum weighted functions. 
These are referred to as transverse moments and typically show up in the description of azimuthal spin asymmetries. 
Describing distribution functions in terms of hadronic
matrix elements, 
the $T$-odd effects come from the gauge-links that 
connect the parton field operators.
The gauge-links are path-ordered exponentials that ensure the 
gauge-invariance of the bilocal products of parton field operators.
In transverse momentum integrated correlators the nonlocality is along 
a lightlike direction. 
The integration path in the gauge-link runs along this lightlike direction
and can be calculated by resumming all exchanges of collinear gluons between the hadronic remnants and the hard part.
In transverse momentum dependent (TMD) correlators the nonlocality is 
not restricted to a lightlike direction, but rather to a light-front.
It turns out that the integration path of the gauge-link between the two parton fields, 
which in this case involves resumming collinear and transverse gluon interactions~\cite{Belitsky:2002sm,Boer:2003cm}, 
depends on the hard partonic subprocess.
In particular it depends on the color-flow through the subprocess.

The time-reversal odd effects for fragmentation
functions do not have such a clear origin, since their description
in terms of matrix elements of parton field operators is more complicated.
In 1993 a mechanism was identified by Collins to generate $T$-odd effects 
in the final-state~\cite{Collins:1992kk}. 
This mechanism originates from final state interactions between an 
outgoing observed hadron and the accompanying jet. 
Due to the explicit appearance of out-states, 
time-reversal symmetry does not constrain the parametrization of the fragmentation correlators. 
Hence, $T$-odd fragmentation effects arise from both final-state interactions {\em and} gauge-links.

Since there are no corresponding effects for the in-states on the distribution side, 
no $T$-odd effects were expected in the distribution
correlators. This expectation was falsified by Brodsky, Hwang and Schmidt,
who could produce leading order single spin asymmetries in a model calculation,
coming from the soft gluon interactions~\cite{Brodsky:2002cx}. 
The effect is identified with the nontrivial 
behavior of the gauge-link under time reversal, 
in particular for the case of TMD correlators where they have both longitudinal and transverse parts.
In the lightcone gauge the connection could be made to the gluonic pole matrix elements in the Qiu-Sterman mechanism~\cite{Boer:2003cm},
which is known to generate single spin asymmetries in the collinear description of hard processes~\cite{Qiu:1998ia}.

In the simplest hadronic scattering processes, such as semi-inclusive 
deep inelastic scattering (SIDIS), Drell-Yan (DY) scattering 
and $e^+e^-$-annihilation,
only a limited number of different gauge-link structures appear.
These are the well-known future and past-pointing Wilson-lines.
They are related by time-reversal and, correspondingly,
the $T$-odd distribution functions appear with opposite sign. 
When going beyond these processes one may encounter more complicated gauge-link structures~\cite{Bomhof:2004aw,Bacchetta:2005rm,Bomhof:2006dp}.
Consequently, the simple sign relation between $T$-odd effects in SIDIS and DY generalizes to color-dependent factors,
called \emph{gluonic pole factors}.
In particular, it is the color-flow of the hard part that determines 
the structure of the gauge-link and, hence,
also the gluonic pole factors~\cite{Bomhof:2006dp}. 
Using these factors one can construct color gauge-invariant weighted sums of Feynman diagrams,
referred to as \emph{gluonic pole cross sections}.
In azimuthal asymmetries the distribution (and fragmentation) functions 
appearing in the weighted correlators are convoluted with the gluonic 
pole cross sections, similar to the way that distribution functions are convoluted with 
partonic cross sections in the spin-averaged case.
For the quark contributions to inclusive back-to-back pion or jet production 
in hadron-hadron scattering 
($p^\uparrow p{\rightarrow}\pi\pi X$, $p^\uparrow p{\rightarrow}j_1j_2X$) 
this was shown in Ref.~\cite{Bacchetta:2005rm}. 
All the gauge-links appearing at tree-level in such processes were calculated 
in~\cite{Bomhof:2006dp}.
Using these results we will give all gluonic pole factors in 
section~\ref{GPMEs} after introducing the general formalism in 
section~\ref{MainText}.  That will allow us to calculate all the gluonic 
pole cross sections that can contribute to azimuthal asymmetries in 
hadron-hadron scattering in section~\ref{GPCSs}.
As an illustration we will explicitly work out the example of back-to-back 
pion production in polarized proton-proton scattering 
($p^\uparrow p{\rightarrow}\pi\pi X$) in section~\ref{example}.
In the appendix we will enumerate all the definitions of the parton correlators.
We have included them because it is needed to uniquely fix the 
definitions of the gluonic pole factors.

\section{Outline of formalism\label{MainText}}

\begin{figure}
\centering
\begin{minipage}{4cm}
\centering
\hspace{1cm}\includegraphics[width=\textwidth]{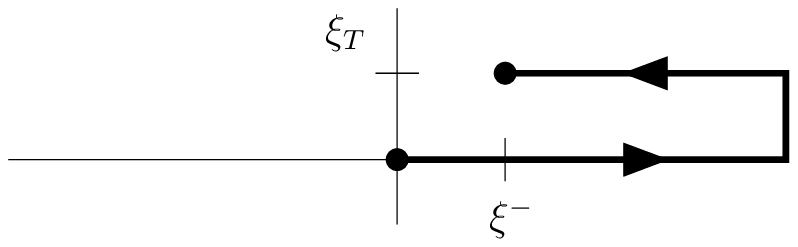}\\[1mm]
(a)
\end{minipage}
\hspace{1.5cm}
\begin{minipage}{4cm}
\centering
\hspace{1cm}\includegraphics[width=\textwidth]{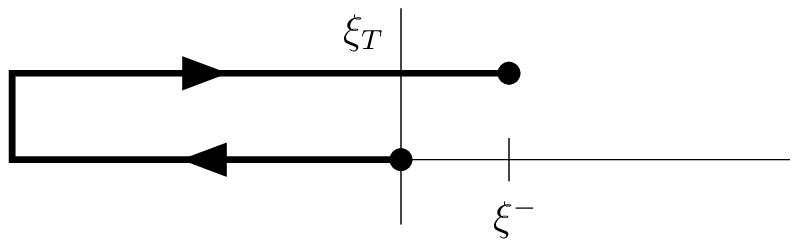}\\[1mm]
(b)
\end{minipage}
\caption{The gauge-link structure in the 
correlator $\Phi$ in (a) SIDIS: $\mathcal U^{[+]}$ and 
(b) DY: $\mathcal U^{[-]}$.\label{simplelinks}}
\end{figure}

We consider processes involving more than one observed hadron, such that
the hadrons are well separated in momentum space. 
This means that the scalar products
of the momenta of any two hadrons must be of the order of the
hard scale squared. In that case the partons entering/exiting the 
hard process are approximately collinear to their parent/daughter hadron
and we can make the following Sudakov decompositions
\begin{subequations}\label{DECOMPS}
\begin{alignat}{2}
p&=xP+\sigma n+p_T\ ,&\qquad
&\text{(incoming particle)}\label{SudakovIn}\\
k&=\frac{1}{z}K+\sigma_hn_h+k_T\ ,&
&\text{(outgoing particle)}\label{SudakovOut}
\end{alignat}
\end{subequations}
where $P$ and $K$ are the momenta of the parent and daughter hadrons, 
respectively, and $p_T{\cdot}P=p_T{\cdot}n=k_T{\cdot}K=k_T{\cdot}n_h=0$.
The $n$ and $n_h$ are arbitrary lightlike vectors.
Note that $p_T$ is transverse to $P$ and $n$,
while $k_T$ is transverse with respect to $K$ and $n_h$.
Using the Sudakov vectors one can define transverse projectors
$g_T^{\mu\nu}{=}g^{\mu\nu}{-}P^{\{\mu}n^{\nu\}}/(P{\cdot}n)$ and 
$\epsilon_T^{\mu\nu}{=}\epsilon^{\mu\nu\rho\sigma}P_\rho n_\sigma/(P{\cdot}n)$ 
for each hadron. The momentum components $x=p{\cdot}n/P{\cdot}n$ and 
$z = K{\cdot}n_h/k{\cdot}n_h$ are of order unity. 
The quantities $\sigma=(p{\cdot}P{-}xM^2)/(P{\cdot}n)$ and $\sigma_h$ 
will appear suppressed by two orders of the hard scale compared to
the leading collinear part $xP$ of the momentum. In the quark
and gluon correlators these momentum components are integrated over,
leaving TMD correlators $\Phi(x,p_T)$ and $\Delta(\frac{1}{z},k_T)$
whose operator definitions are given in appendix~\ref{CORRS},
Eqs.~\eqref{QUARKCORR}-\eqref{GLUONCORR}.
They are parameterized in terms of distribution functions and fragmentation
functions, respectively. 
Regarding color, the quark and antiquark fields belong to the fundamental 
representation. For the gluon field strength we use the matrix representation 
$F_{\mu\nu}{=}F_{\mu\nu}^at^a$, where the $t^a$ are 
the color matrices in the fundamental representation with normalization
$\tr[t^at^b]{=}T_F\delta^{ab}$. 
Also the gauge-links 
$\mathcal U^{\mathcal C}
{=}\mathcal P\exp[{-}ig\int_{\mathcal C}dz{\cdot}A^at^a]$ 
are three-dimensional matrices in color-space.
In TMD distribution and fragmentation functions the field operators 
in the hadronic matrix elements are separated in the transverse and in 
the light-cone $n$-direction~\cite{Belitsky:2002sm,Boer:2003cm}.
In that case there is no unique gauge-link to connect the partonic field 
operators.
For instance, in the quark correlator in SIDIS one has the future pointing 
Wilson line $\mathcal U^{[+]}$ (Fig.~\ref{simplelinks}a),
while in DY one has the past pointing Wilson line 
$\mathcal U^{[-]}$ (Fig.~\ref{simplelinks}b).
When integrating over all intrinsic transverse momenta to obtain the 
collinear correlators,
\begin{subequations}
\begin{alignat}{1}
&\Phi(x) =\int d^2p_T\ \Phi^{[\mathcal U]}(x,p_T)\ , \\
&\Delta\big(\tfrac{1}{z}\big)
=z^2\int d^2k_T\ \Delta^{[\mathcal U]}\big(\tfrac{1}{z},k_T\big)\ ,
\label{kaassoufle-a}
\end{alignat}
\end{subequations}
the gauge-link structure reduces to a simple Wilson line along the light-cone $n$ axis:
\begin{equation}\label{WILSONLINES}
\mathcal U_{[0;\xi]}^{\mathcal C}
\rightarrow 
U^n_{[0;\xi]}=\mathcal P\exp\Big[-ig\int_0^\xi dz\ n{\cdot}A(z)\,\Big]\ .
\end{equation}
This removes all process dependence of the gauge-links in collinear correlators.
In the weighted correlators (transverse moments)
\begin{subequations}\label{weighed}
\begin{gather}
\Phi_\partial^{[\mathcal U]\alpha}(x)
=\int d^2p_T\ p_T^\alpha\ \Phi^{[\mathcal U]}(x,p_T)\ ,\\
\Delta_\partial^{[\mathcal U]\alpha}\big(\tfrac{1}{z}\big)
=z^2\int d^2k_T\ k_T^\alpha\ \Delta^{[\mathcal U]}\big(\tfrac{1}{z},k_T\big)\ ,
\end{gather}
\end{subequations}
a gauge-link dependence remains, 
which (as will be discussed below) can be expressed as a gluonic pole matrix
element multiplied by prefactors that depend on the particular
path structure of the gauge-link.
A well known example is the Sivers effect in
SIDIS and DY~\cite{Collins:2002kn,Brodsky:2002rv}. The 
weighted distribution correlator contains among others the $T$-odd 
Sivers function $f_{1T}^{\perp(1)}(x)$ which, 
depending on the integration path of the gauge-link, appears with
a plus or minus sign in the cross section,
\begin{subequations}\label{SiversEffect}
\begin{alignat}{3}
&\text{Sivers effect in SIDIS:}&\qquad
&d\sigma_{\ell H{\rightarrow}\ell h X}&&\sim \ 
+f_{1T}^{\perp(1)}(x)
\,d\hat\sigma_{\ell q{\rightarrow}\ell q}\,D_1(z)\ ,\\
&\text{Sivers effect in DY:}&
&d\sigma_{HH^\prime{\rightarrow}\ell\bar\ell X}&\ &\sim \ 
-f_{1T}^{\perp(1)}(x)\,\bar f_1(x')
\,d\hat\sigma_{q\bar q{\rightarrow}\ell\bar\ell}\ .
\end{alignat}
\end{subequations}
For other processes the plus and minus signs generalize to other factors.
Furthermore, the prefactor may differ for each Feynman diagram $D$ that 
contributes to the partonic scattering cross section.
Therefore, single spin asymmetries can, in general, 
no longer be written as convolutions of distribution and fragmentation 
functions with the standard partonic cross sections
\begin{equation}\label{PARTONCROSSSECTION}
\frac{d\hat\sigma_{ab\rightarrow cd}}{d\hat t}
=\sum_D\,\frac{d\hat\sigma^{[D]}_{ab\rightarrow cd}}{d\hat t}\ ,
\end{equation}
(the usual minus signs in front of diagrams that are related to each other by interchanging two identical fermions in the initial/final state or by moving a fermion from the initial to the final state and vice versa are included in this summation).
In~\cite{Bacchetta:2005rm} it was shown that in some cases one can still cast 
the expression for the single spin asymmetries in the form of such a 
convolution provided that one introduces modified hard parts. 
These modified hard parts, referred to as \emph{gluonic pole cross sections},
are the sum of the different terms in the squared amplitude 
(if more than one diagram contributes) multiplied with the respective 
prefactors that arise from the gauge-link,
\begin{equation}\label{GPCS}
\frac{d\hat\sigma_{[a]b\rightarrow cd}}{d\hat t}
=\sum_D\,C_G^{[D]}\,
\frac{d\hat\sigma^{[D]}_{ab\rightarrow cd}}{d\hat t}\ .
\end{equation}
In this example it is parton $a$ that contributes the $T$-odd distribution 
function, indicated by the bracketed subscript $[a]$ on the gluonic pole 
cross section.
If it is another parton that contributes the $T$-odd function we use a similar notation for its corresponding subscript.
In the simplest of all hadronic scattering processes, SIDIS and DY scattering,
the hard processes consist of single diagrams (at leading order).
Therefore, the partonic and the gluonic pole cross sections are simply proportional to each other in these processes,
with proportionality factors $C_G^{[\pm]}{=}\pm$ 
(cf.~equation~\eqref{SiversEffect}),
and it seems superfluous to define the gluonic pole cross sections at all.
However, going beyond SIDIS and DY there are, in general, 
several diagrams that contribute to the hard partonic scattering
(we will give an example in section~\ref{example}).
The partonic and gluonic pole cross sections are, then, 
no longer necessarily proportional to each other.
In that case the use of gluonic pole cross sections comes as a natural definition since, as was already mentioned earlier, 
the gluonic pole strengths $C_G^{[D]}$ depend solely on the structure of the gauge-link which, in turn,
is fixed by the color-flow structure of the Feynman diagram $D$~\cite{Bacchetta:2005rm,Bomhof:2006dp}.
In what follows we will summarize how they can be calculated.
Using the results of~\cite{Bomhof:2006dp} we will give all gluonic 
pole factors and the corresponding gluonic pole cross sections that can appear in the tree-level contributions to single spin asymmetries in hadron-hadron scattering.

The expressions $d\hat\sigma^{[D]}$ that appear in the sums~\eqref{PARTONCROSSSECTION} and~\eqref{GPCS} are bilinear combinations of amplitudes.
More precisely, for unpolarized scattering they are defined through
$d\hat\sigma^{[D]}_{ab\rightarrow cd}
{=}\frac{1}{16\pi\hat s^2}\frac{1}{4}
M_{ab\rightarrow cd}^{\phantom{*}}M_{ab\rightarrow cd}^*$,
where $M$ is the amplitude and $M^*$ the conjugate amplitude,
whose product is pictorially represented by the Feynman diagram $D$.
For polarized scattering we use the expressions defined in~\cite{Stratmann:1992gu,Vogelsang:1998yd}:
\begin{equation*}
d\Delta\hat\sigma^{[D]}_{a^\uparrow b\rightarrow c^\uparrow d}
=\frac{1}{16\pi\hat s^2}\frac{1}{4}
\big(\,M_{a^\uparrow b\rightarrow c^\uparrow d}^{\phantom{*}}
M_{a^\uparrow b\rightarrow c^\uparrow d}^*
-M_{a^\uparrow b\rightarrow c^\downarrow d}^{\phantom{*}}
M_{a^\uparrow b\rightarrow c^\downarrow d}^*\,\big)\ ,
\end{equation*}
and
\begin{equation*}
d\Delta\hat\sigma^{[D]}_{a_Lb\rightarrow c_Ld}
=\frac{1}{16\pi\hat s^2}\frac{1}{4}
\big(\,M_{a^+b\rightarrow c^+d}^{\phantom{*}}
M_{a^+b\rightarrow c^+d}^*
-M_{a^+b\rightarrow c^-d}^{\phantom{*}}
M_{a^+b\rightarrow c^-d}^*\,\big)\ ,
\end{equation*}
where ${\uparrow}{\downarrow}$ and $\pm$ refers to the transverse spin and helicity of the associated particles, respectively.
For unpolarized particles a summation over spins is understood.
An averaging over the color quantum numbers of the incoming particles is also implied. 
These expressions are, themselves, not gauge-invariant.
However, the sum of Feynman diagrams~\eqref{PARTONCROSSSECTION} and the weighted sum of Feynman diagrams~\eqref{GPCS} are properly gauge-invariant.

\section{Gluonic Pole Matrix Elements\label{GPMEs}}

In the calculation of physical observables, the full transverse momentum 
dependence requires the inclusion of many correlators.
This includes besides TMD quark-quark or gluon-gluon correlators, 
also correlators with additional gluons.
These give rise to the gauge-links in the gauge-invariant correlator
$\Phi^{[\mathcal U]}(x,p_T)$.
To display the connection between the gauge-links and the $T$-odd effects
we first consider the situation in SIDIS and DY.
In Ref.~\cite{Boer:2003cm} it was explicitly shown how the weighted correlators 
with future/past pointing Wilson lines (Figure~\ref{simplelinks}) can be 
decomposed into two parts,
\begin{align}
\Phi^{[\pm]\alpha}_\partial(x)
=\widetilde\Phi_\partial^\alpha(x)
\pm\pi\Phi_G^\alpha(x,x)\ ,
\label{SIDISDYcorrB}
\end{align}
with the matrix element $\widetilde\Phi_\partial$ and the gluonic
pole matrix element $\Phi_G$ defined in appendix~\ref{CORRS}.
The decomposition in~\eqref{SIDISDYcorrB} is particularly useful
since $\widetilde\Phi_\partial(x)$ and $\Phi_G(x,x)$ have opposite 
time reversal behavior, i.e., 
$\widetilde\Phi_\partial(x)$ is $T$-even and $\Phi_G(x,x)$ is $T$-odd.
Hence, for distribution functions the gluonic pole matrix element is
the sole source for the $T$-odd distribution functions.
For instance, the Sivers function $f_{1T}^{\perp(1)}(x)$ is proportional 
to the projection $\tr[\,\slash n\,\pi\Phi_G(x,x)\,]$.
The relative minus sign in the Sivers effect in SIDIS and 
DY~\eqref{SiversEffect} originates from the different (link-dependent)
signs in the decomposition~\eqref{SIDISDYcorrB}.

In other processes one may encounter distribution correlators with more 
complicated gauge-links than the simple future or past pointing Wilson lines.
In those cases one can still make a decomposition such as 
in~\eqref{SIDISDYcorrB}, 
but with different factors in front of the gluonic pole matrix element:
\begin{equation}
\Phi^{[\mathcal U]\alpha}_\partial(x)
=\widetilde\Phi_\partial^\alpha(x)
+C_G^{[\mathcal U]}\,\pi\Phi_G^\alpha(x,x)\ .
\end{equation}
This implies that the $T$-odd functions in the parametrization of the
transverse moments, which originate from the gluonic pole part $\Phi_G(x,x)$,
appear with these prefactors $C_G^{[\mathcal U]}$. 
Since these factors multiply the gluonic pole matrix elements they are referred to as \emph{gluonic pole strengths}.
They can be different for each Feynman diagram $D$ that contributes 
to a particular process and can be calculated by taking the first moment
of the TMD correlators containing the process-dependent gauge-links 
$\mathcal U{=}\mathcal U(D)$, see equation~\eqref{weighed}.
In Ref.~\cite{Bacchetta:2005rm} all gluonic pole strengths in $2{\rightarrow}2$ (anti)quark scattering processes were calculated in this way.
For completeness we reproduce the results here in the Tables~\ref{Tqq2qq} and~\ref{Tqq_2qq_}.
As the gluonic pole factors for a given gauge-link only depend on the color structure of the Feynman diagram,
they are gauge-invariant quantities.

In fragmentation the discussion is slightly more complicated,
since the gauge-links are not the only source of $T$-odd effects.
As pointed out by Collins, also the \emph{internal} final state 
interactions of the observed outgoing hadron with its accompanying 
jet, in matrix elements appearing as the one-particle inclusive out-state 
$\vert h{,}X\rangle$,
can produce $T$-odd effects~\cite{Collins:1992kk}.
In Ref.~\cite{Boer:2003cm} it was shown that the first moment of the quark 
fragmentation correlator with future ($e^+e^-$) and past (SIDIS) pointing 
Wilson lines can be decomposed as follows:
\begin{equation}\label{SIDISDYfragcorr}
\Delta^{[\pm]\alpha}_\partial\big(\tfrac{1}{z}\big)
=\widetilde\Delta_\partial^\alpha\big(\tfrac{1}{z}\big)
\pm\pi\Delta_G^\alpha\big(\tfrac{1}{z},\tfrac{1}{z}\big)\ .
\end{equation}
The matrix elements appearing in this expression are defined in 
appendix~\ref{CORRS}.
In other processes one may again encounter fragmentation correlators with more 
complicated gauge-links than the simple future or past pointing Wilson lines.
In those cases one can also make a decomposition such as described above, 
but with different factors in front of the gluonic pole matrix element:
\begin{equation}\label{GPSqqFRAG}
\Delta^{[\mathcal U]\alpha}_\partial\big(\tfrac{1}{z}\big)
=\widetilde\Delta_\partial^\alpha\big(\tfrac{1}{z}\big)
+C_G^{[\mathcal U]}\,
\pi\Delta_G^\alpha\big(\tfrac{1}{z},\tfrac{1}{z}\big)\ .
\end{equation}
Due to the presence of out-states in the matrix elements
$\widetilde\Delta_\partial$ and $\pi\Delta_G$, 
both contain $T$-even {\em and}
$T$-odd fragmentation functions.
The parametrization of both these matrix elements contain, for instance,
a Collins-effect-like fragmentation function.
For definiteness we use $H_1^{\perp(1)}(z)$ to refer to the one that 
appears in the parametrization of $\widetilde\Delta_\partial(\tfrac{1}{z})$ 
and $\widetilde H_1^{\perp(1)}(z)$ for the function in the gluonic pole matrix element
$\pi\Delta_G(\frac{1}{z},\frac{1}{z})$~\cite{Bacchetta:2005rm}.
Such tilde-fragmentation functions find their origin in the gauge-links,
while the others find their origin in the final-state interactions
included in the out-states.  
Consequently, from expression~\eqref{SIDISDYfragcorr} it follows that in SIDIS the Collins effect is given by
$H_1^{\perp(1)}{-}\widetilde H_1^{\perp(1)}$,
while in electron-positron annihilation it is given by 
$H_1^{\perp(1)}{+}\widetilde H_1^{\perp(1)}$.
In Ref.~\cite{Collins:2004nx} it is argued that the Collins effect in these 
two processes is described by a single universal function.
In this paper we will take the general approach in which all gluonic pole matrix elements are included.
The universality as advocated in Ref.~\cite{Collins:2004nx} is the limiting case of vanishing gluonic pole matrix elements 
$\pi\Delta_G(\frac{1}{z},\frac{1}{z})$ in fragmentation.

\begin{figure}
\includegraphics[width=3.6cm]{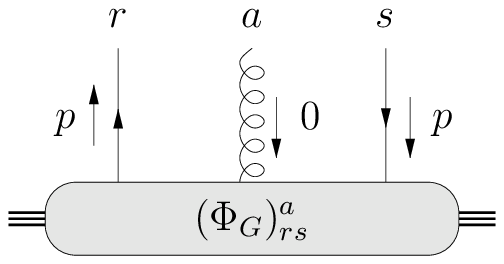}\hspace{1cm}
\includegraphics[width=3.6cm]{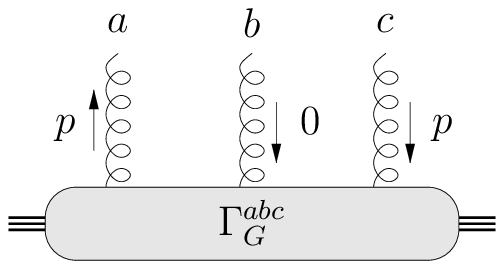}
\parbox{0.89\textwidth}{
\caption{We encounter three different types of gluonic pole matrix elements (GPME's):
the quark-GPME $\Phi_G{=}(\Phi_G)_{rs}^a(t^a)_{sr}{=}\tr[\Phi_G^at^a]$
and the two gluon-GPME's 
$\Gamma_G^{(f)}{=}\Gamma_G^{abc}(t^a)_{cb}{=}\Gamma_G^{abc}if^{abc}$ and $\Gamma_G^{(d)}{=}\Gamma_G^{abc}d^{abc}$.
The $d^{abc}$ and $f^{abc}$ are the symmetric and antisymmetric 
structure constants of $SU(3)$:
$t^at^b{=}\frac{T_F}{N}\delta^{ab}
{+}\frac{1}{2}\left(d^{abc}{+}if^{abc}\right)t^c$.
\label{GLUONgluonicPOLES}}}
\end{figure}

The treatment of antiquark distribution and fragmentation follows along 
similar lines as described above.
The results are summarized in appendix~\ref{CORRS}.
Gluon distribution correlators are written as the product of two gluon 
field-strength tensors and two gauge-links connecting them, see~\eqref{GLUONCORR}.
The product of operators is in the 3-dimensional fundamental representation 
of the color matrices and appears traced in the correlators.
Due to the presence of two gauge-links (in the matrix representation),
rather than one, as was the case in quark distribution,
there are two different types of gluonic pole matrix elements:
those with soft gluons arising from the gauge-link $\mathcal U$ or $\mathcal U^\prime$, respectively. 
This complicates the occurrence of $T$-odd effects in gluon
distributions somewhat.
Since the correlator is color-traced, 
it is convenient to consider the commutator and anticommutator combinations of gluon fields in the matrix elements.
These involve the symmetric and the antisymmetric
structure constants of $SU(3)$ color (Figure~\ref{GLUONgluonicPOLES}). 
Taking the first moment of the gluon correlator~\eqref{GLUONCORRa}
it follows that
\begin{equation}\label{SINTERKLAAS}
\Gamma^{[\mathcal U\mathcal U^\prime]\alpha}_\partial(x)
=\widetilde\Gamma_\partial^\alpha(x)
+C_G^{(f)}\,\pi\Gamma_G^{(f)}{}^{\alpha}(x,x)
+C_G^{(d)}\,\pi\Gamma_G^{(d)}{}^{\alpha}(x,x)\ ,
\end{equation}
where the gluonic pole strengths $C_G^{(f/d)}$ involve sums and differences of contributions arising from the two different gauge-links. 
The matrix element $\widetilde\Gamma_\partial$ 
(which is purely $T$-even in the case of gluon distribution)
only involves a color commutator 
(cf.~\eqref{appelsap} and~\eqref{sinaasappelsap}).
The definitions of the different matrix elements appearing in this 
expression are given in appendix~\ref{CORRS}.
The leading contributions to the collinear gluon correlator is parametrized 
as follows
\begin{equation}\label{GLUONparametrization}
2x\,\Gamma^{ij}(x)
=-g_T^{ij}\,G(x)
-S_L\,i\epsilon_T^{ij}\,\Delta G(x)\ .\\
\end{equation}
For the gluon fragmentation correlator we use a similar parametrization.
It is obtained by making the substitutions
$\big\{2x\Gamma(x),G(x),\Delta G(x)\big\}{\rightarrow}
\big\{\widehat\Gamma(\frac{1}{z}),\widehat G(z),\Delta\widehat G(z)\big\}$.
Following the treatment of TMD functions in Ref.~\cite{Mulders:2000sh}
we take for the $\widetilde\Gamma_\partial(x)$ and the gluonic pole matrix 
elements the parametrizations
\begin{subequations}\label{GPMelement}
\begin{gather}
2x\,\widetilde\Gamma^{ij;\alpha}(x)
=-M\,i\epsilon_T^{ij}S_T^\alpha\,\Delta G_T^{(1)}(x)\ ,\\
2x\,\pi\Gamma_G^{(f/d)}{}^{ij;\alpha}(x,x)
=M\,g_T^{ij}\epsilon_T^{S_T\alpha}\,G_T^{(f/d)}{}^{(1)}(x)
-\tfrac{1}{2}M\,\big(\epsilon_T^{\alpha\{i}S_T^{j\}}{+}
\epsilon_T^{S_T\{i}g_T^{j\}\alpha}\big)\,\Delta H_T^{(f/d)}{}^{(1)}(x)\ .
\label{GPMelement1}
\end{gather}
\end{subequations}
The indices $i$, $j$ and $\alpha$ are all transverse w.r.t.\ the hadron 
direction and its conjugate direction (cf.~\eqref{DECOMPS}).
The functions $\Delta H_T^{(1)}$ involve a flip of gluon helicity in a 
transversally polarized target.
The functions $G_T^{(1)}$ and $\Delta G_T^{(1)}$ do not flip the gluons 
helicity and represent unpolarized and polarized gluons in transversally 
polarized targets, respectively.
Note that the $T$-odd quark and gluon distribution (fragmentation) functions that appear in the parametrizations of the gluonic pole matrix elements are different from quark-quark or gluon-gluon matrix elements.
They contain an additional zero-momentum gluon (the gluonic pole).
Therefore, it is not obvious if such $T$-odd functions can be interpreted as probability distributions.
The emergence of two (the $f$ and the $d$) $T$-odd gluon distribution (or fragmentation) functions is simply the consequence of the two possible orderings of the three gluons in the matrix element $\Gamma_G$ in figure~\ref{GLUONgluonicPOLES}.

From symmetry relations it follows that $G_T^{(f/d)}{}^{(1)}(x)$ 
can be written as the $x$-even and $x$-odd combinations of a single gluon 
distribution function, i.e.\ $G_T^{(f/d)}{}^{(1)}(x)
{=}\frac{1}{2}\big(G_T^{(1)}(x){\pm}G_T^{(1)}({-}x)\big)$ and
similarly for $\Delta H_T^{(f/d)}{}^{(1)}(x)$.
As seen in equation~\eqref{SINTERKLAAS} the $x$-even and $x$-odd combinations 
of gluon distribution functions can appear with different prefactors.
Hence, the description of the gluon contribution to $T$-odd effects involves 
two different gluonic pole cross sections,
one containing the gluonic pole factors $C_G^{(d)}$ and one containing 
the factors $C_G^{(f)}$.
Using the results of~\cite{Bomhof:2006dp} we can give all gluonic pole 
strengths $C_G$ appearing in $2{\rightarrow}2$ scattering processes 
with external gluons. They are listed in Tables~\ref{Tqq_2gg}-\ref{Tgg2gg}.
The treatment of gluon fragmentation is the obvious extension of the 
formalism described above.

The gluonic pole strengths associated with quarks $C_G(q)$ calculated
in the way described above are related to the color factors calculated by
Qiu and Sterman in Ref.~\cite{Qiu:1998ia}, who 
consider one-pion inclusive proton-proton scattering
$pp{\rightarrow}\pi X$. The relation is~\cite{Vogelsang}
\begin{equation}\label{QSfactors}
C_G(q)=\frac{C_{q/g}^I{+}C_{q/g}^F{+}C_{q/g}^{F'}}{C_{q/g}}\ ,
\end{equation}
where $C_{q/g}$, $C_{q/g}^I$ and $C_{q/g}^F$ are the color factors
defined and calculated in Ref.~\cite{Qiu:1998ia}
(one should note that some of the color factors in Table~II of that reference are erroneous).
The factor $C_{q/g}$ is the color factor of the spin-averaged subprocess.
The factor $C_{q/g}^I$ is the color factor of the spin-dependent
subprocess with a gluon attachment on the (other) incoming parton and
the factor $C_{q/g}^F$ is the color factor of the spin-dependent 
subprocess with a gluon attachment on the outgoing parton that fragments 
into the observed pion. 
The factor $C_{q/g}^{F'}$ is the corresponding color factor of the spin-dependent subprocess with a gluon attachment on the other outgoing parton.
These factors have not been given in Ref.~\cite{Qiu:1998ia},
where it is argued that they are not needed in the one-pion
inclusive process.
They can be calculated in exactly the same way as the factors $C^I$ and
$C^F$.
They can also be obtained from symmetry considerations.
The color factor $C^{F'}$ of the process $ab{\rightarrow}c_\pi d$ should be the
same as the color factor $C^F$ of the process $ab{\rightarrow}cd_\pi$
(the subscript $\pi$ indicating which parton fragments into the observed pion).

\begin{figure}
\centering
\begin{minipage}[t]{2.5cm}
\includegraphics[width=2.5cm]{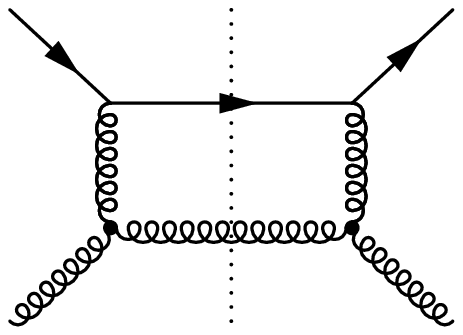}\\[1mm]
(a)
\end{minipage}\hspace{3mm}
\begin{minipage}[t]{2.5cm}
\includegraphics[width=2.5cm]{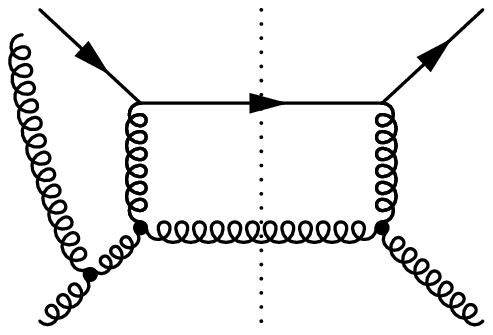}\\[1mm]
(b)
\end{minipage}\hspace{3mm}
\begin{minipage}[t]{2.5cm}
\includegraphics[width=2.5cm]{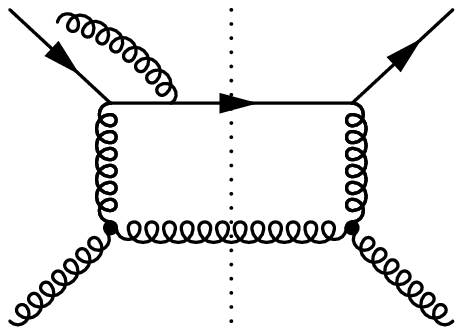}\\[1mm]
(c)
\end{minipage}\hspace{3mm}
\begin{minipage}[t]{2.5cm}
\includegraphics[width=2.5cm]{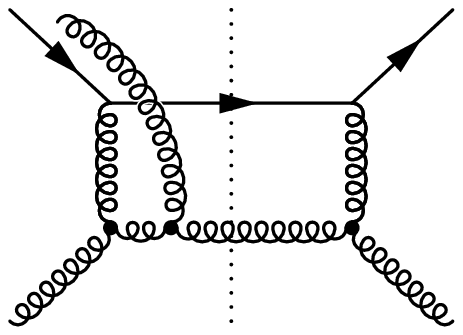}\\[1mm]
(d)
\end{minipage}
\caption{A contribution to $qg{\rightarrow}qg$ scattering (a)
and some collinear gluon insertions (b)-(d).\label{QIUSTERMAN}}
\end{figure}

As an example we consider the contribution of the diagram in
Figure~\ref{QIUSTERMAN}a to quark-gluon scattering $qg{\rightarrow}qg$ 
explicitly.
In this example the quark comes from the polarized hadron and it is also
the quark that fragments into the observed pion.
In that case Figures~\ref{QIUSTERMAN}a,~\ref{QIUSTERMAN}b
and~\ref{QIUSTERMAN}c
correspond to the Figures~15a,~16a and~16b of Ref.~\cite{Qiu:1998ia}. 
From the first line of Table~I in~\cite{Qiu:1998ia} one finds their
color factors: 
$C_g{=}\frac{1}{2}$, $C_g^I{=}{-}N^2/4(N^2{-}1)$ and
$C_g^F{=}{-}1/2(N^2{-}1)$.
The color factor corresponding to the gluon insertion in
Figure~\ref{QIUSTERMAN}d is calculated to be $C_g^{F'}{=}N^2/4(N^2{-}1)$.
Hence, the relation in Eq.~\eqref{QSfactors} indeed reproduces 
the gluonic pole strength $C_G(q){=}{-}1/(N^2{-}1)$.
The color factor $C_g^{F'}$ that we have just calculated would have been
the same as the color factor $C_g^F$ of the corresponding contribution to
$qg{\rightarrow}qg_\pi$ scattering if Ref.~\cite{Qiu:1998ia} would have 
considered the possibility that it is the gluon that fragments into the observed pion.

\section{Gluonic Pole Scattering Cross Sections\label{GPCSs}}

\begin{figure}
\centering
\begin{minipage}{6cm}
\centering
\includegraphics[width=6cm]{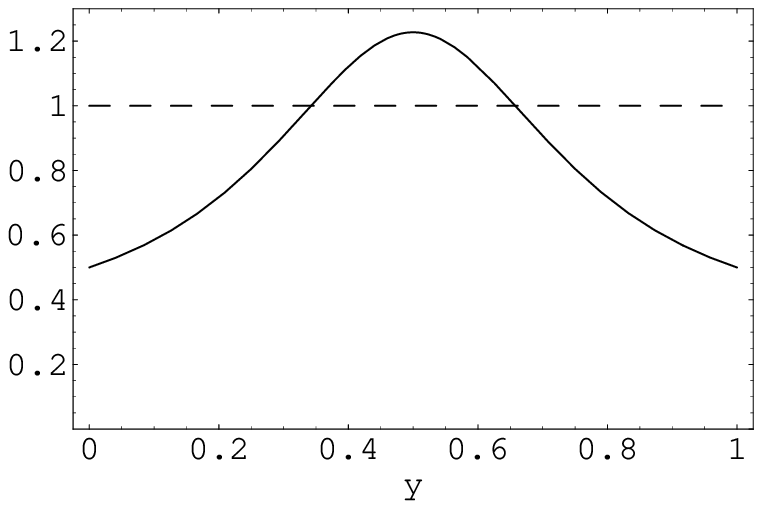}\\
\hspace{2mm}
$d\hat\sigma_{[q]q\rightarrow qq}/d\hat\sigma_{qq\rightarrow qq}$
\end{minipage}\hspace{1.5cm}
\begin{minipage}{6cm}
\centering
\includegraphics[width=6cm]{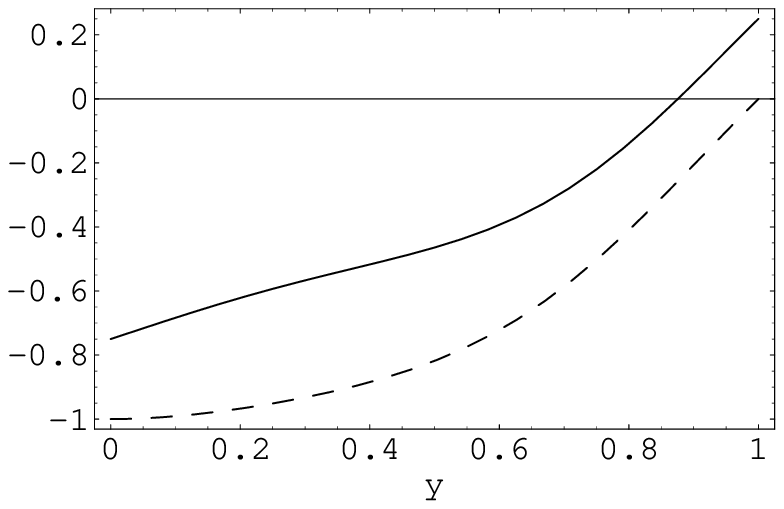}\\
\hspace{2mm}
$d\hat\sigma_{[q]\bar q\rightarrow q\bar q}/
d\hat\sigma_{q\bar q\rightarrow q\bar q}$
\end{minipage}\\[5mm]
\begin{minipage}{6cm}
\centering
\includegraphics[width=6cm]{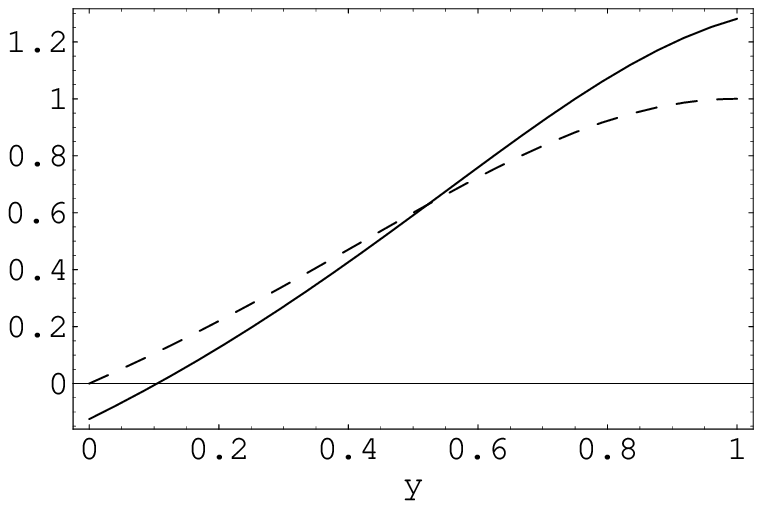}\\
\hspace{2mm}
$d\hat\sigma_{[q]g\rightarrow qg}/d\hat\sigma_{qg\rightarrow qg}$
\end{minipage}\hspace{1.5cm}
\begin{minipage}{6cm}
\centering
\includegraphics[width=6cm]{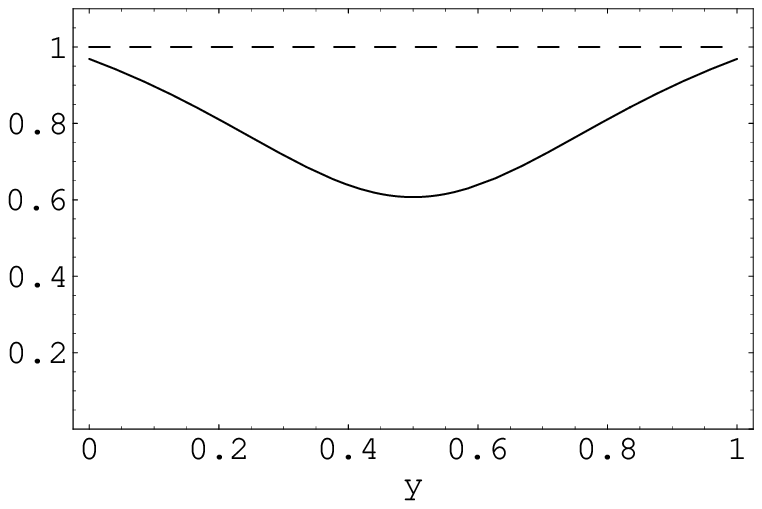}\\
\hspace{2mm}
$d\hat\sigma_{[q]\bar q\rightarrow gg}/d\hat\sigma_{q\bar q\rightarrow gg}$
\end{minipage}
\parbox{0.92\textwidth}{
\caption{Ratios of gluonic pole cross sections and partonic cross sections for some gluonic pole cross sections associated with quarks.
They are plotted as functions of the variable $y{\equiv}{-}\hat t/\hat s$ 
for $N{=}3$ (solid line) and $N{\rightarrow}\infty$ (dashed line).
\label{RelDifferences}}}
\end{figure}

As indicated in previous sections we aim at a description of single spin asymmetries as convolutions of universal collinear parton 
distribution and fragmentation functions and perturbatively calculable 
hard parts. These convolutions involve an odd number of (usually one) $T$-odd 
distribution or fragmentation functions.
The basic observation of the previous section is that the gluonic pole
matrix elements that contain the $T$-odd functions are 
multiplied by gluonic pole factors $C_G$. 
These might be different for each Feynman diagram $D$ that contributes to the partonic scattering process.
Consequently, the hard part in the description of the SSA, in general, 
no longer equals the basic partonic scattering cross 
section in Eq.~\eqref{PARTONCROSSSECTION},
but rather the gluonic pole cross sections in Eq.~\eqref{GPCS}.
For instance, in the identical quark scattering contribution to 
$p^\uparrow p{\rightarrow}\pi\pi X$  (see next section)
the direct scattering channels 
(first and second diagram in Table~\ref{Tqq2qq}) 
get the factor $C_G{=}(N^2{-}5)/(N^2{-}1){=}\frac{1}{2}$ 
and the interference terms (third and fourth diagram in Table~\ref{Tqq2qq}) 
get the factor $C_G{=}{-}(N^2{+}3)/(N^2{-}1){=}{-}\frac{3}{2}$ from the 
incoming quark.  Accordingly, the Sivers effect for the identical 
quark scattering contribution to $p^\uparrow p{\rightarrow}\pi\pi X$ appears
proportional to
\begin{equation}
f_{1T}^{\perp(1)}(x_1)\,f_1(x_2)\,\bigg(
\{\tfrac{1}{2}\}
\parbox{1.5cm}{\includegraphics[width=1.5cm]{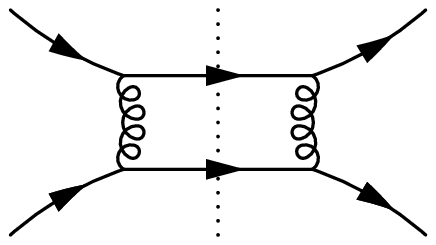}}
+\{\tfrac{1}{2}\}
\parbox{1.5cm}{\includegraphics[width=1.5cm]{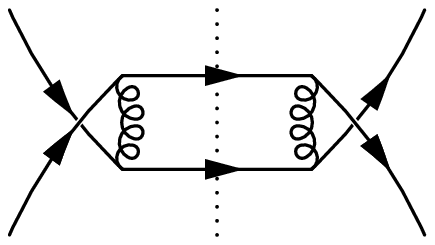}}
-\{{-}\tfrac{3}{2}\}
\parbox{1.5cm}{\includegraphics[width=1.5cm]{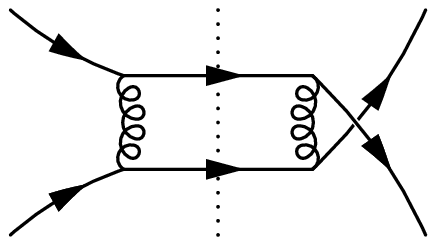}}
-\{{-}\tfrac{3}{2}\}
\parbox{1.5cm}{\includegraphics[width=1.5cm]{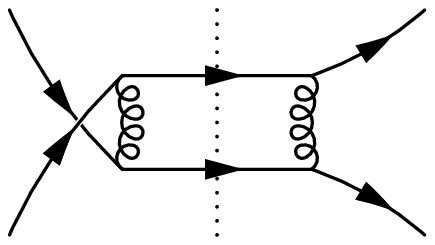}}\,\bigg)\,
D_1(z_1)\,D_1(z_2)\ ,
\end{equation}
(compare this with the situation in SIDIS and DY, cf.\ Eq.~\eqref{SiversEffect}).
Therefore, as a consequence of the gauge-links the contribution of the 
Sivers effect is a convolution of the Sivers function with the unpolarized 
distribution and fragmentation functions and the hard cross section
\begin{equation}
\frac{d\hat\sigma_{[q]q{\rightarrow}qq}}{d\hat t}
=\tfrac{1}{2}
\parbox{1.5cm}{\includegraphics[width=1.5cm]{Figures/qq2qqA.eps}}
+\tfrac{1}{2}
\parbox{1.5cm}{\includegraphics[width=1.5cm]{Figures/qq2qqB.eps}}
+\tfrac{3}{2}
\parbox{1.5cm}{\includegraphics[width=1.5cm]{Figures/qq2qqC.eps}}
+\tfrac{3}{2}
\parbox{1.5cm}{\includegraphics[width=1.5cm]{Figures/qq2qqD.eps}}\,
=\frac{4\pi\alpha_S^2}{9\hat s^2}\,\bigg\{\,
\frac{\hat s^2{+}\hat u^2}{2\hat t^2}+\frac{\hat s^2{+}\hat t^2}{2\hat u^2}
+\frac{\hat s^2}{\hat t\hat u}\,\bigg\}\ .
\end{equation}
We refer to this as a gluonic pole scattering cross section 
and it should be contrasted to the standard partonic cross section 
\begin{equation}
\frac{d\hat\sigma_{qq{\rightarrow}qq}}{d\hat t}
=\parbox{1.5cm}{\includegraphics[width=1.5cm]{Figures/qq2qqA.eps}}
+\parbox{1.5cm}{\includegraphics[width=1.5cm]{Figures/qq2qqB.eps}}
-\parbox{1.5cm}{\includegraphics[width=1.5cm]{Figures/qq2qqC.eps}}
-\parbox{1.5cm}{\includegraphics[width=1.5cm]{Figures/qq2qqD.eps}}\,
=\frac{4\pi\alpha_S^2}{9\hat s^2}\,\bigg\{\,
\frac{\hat s^2{+}\hat u^2}{\hat t^2}+\frac{\hat s^2{+}\hat t^2}{\hat u^2}
-\frac{2}{3}\frac{\hat s^2}{\hat t\hat u}\,\bigg\}\ ,
\end{equation}
(see Figure~\ref{RelDifferences}).

For quark fragmentation the situation is more involved.
As mentioned, due to the internal final state interactions
both terms on the r.h.s.\ of expression~\eqref{GPSqqFRAG} can contribute to the $T$-odd effects.
The matrix element $\widetilde\Delta_\partial\big(\tfrac{1}{z}\big)$ appears 
in the same way, regardless of the process, while the gluonic pole matrix 
element occurs multiplied by the gluonic pole factors $C_G$, which 
depend on the gauge-links and, hence, on the hard subprocess.
Comparing this with the case of quark distribution described above,
we see that the non-tilde-functions appear with the standard partonic cross sections and the tilde-functions appear with gluonic pole cross sections.
For instance, for the identical quark subprocess in 
$p^\uparrow p{\rightarrow}\pi\pi X$ scattering the gluonic pole factors associated with the fragmenting quark are $C_G{=}{-}\frac{1}{2}$ for the direct scattering channels and $C_G{=}\frac{3}{2}$ for the interference terms.
Therefore, the Collins effect becomes
\begin{equation}
h_1(x_1)\,f_1(x_2)\,
\frac{d\Delta\hat\sigma_{q^\uparrow q{\rightarrow}q^\uparrow q}}{d\hat t}\,
H_1^{\perp(1)}(z_1)\,D_1(z_2)
+h_1(x_1)\,f_1(x_2)\,
\frac{d\Delta\hat\sigma_{q^\uparrow q{\rightarrow}
[q^\uparrow] q}}{d\hat t}\,
\widetilde H{}_1^{\perp(1)}(z_1)\,D_1(z_2)\ ,
\end{equation}
where $h_1$ is the transversity distribution function.
The $d\Delta\hat\sigma_{q^\uparrow q{\rightarrow}q^\uparrow q}/d\hat t$
is the standard partonic cross section and 
$d\Delta\hat\sigma_{q^\uparrow q{\rightarrow}[q^\uparrow] q}/d\hat t$
is the gluonic pole cross section
\begin{subequations}
\begin{gather}
\frac{d\Delta\hat\sigma_{q^\uparrow q{\rightarrow}q^\uparrow q}}{d\hat t}
=\parbox{1.5cm}{\includegraphics[width=1.5cm]{Figures/qq2qqA.eps}}
+\parbox{1.5cm}{\includegraphics[width=1.5cm]{Figures/qq2qqB.eps}}
-\parbox{1.5cm}{\includegraphics[width=1.5cm]{Figures/qq2qqC.eps}}
-\parbox{1.5cm}{\includegraphics[width=1.5cm]{Figures/qq2qqD.eps}}\ ,\\
\frac{d\Delta\hat\sigma_{q^\uparrow q{\rightarrow}
[q^\uparrow] q}}{d\hat t}
=-\tfrac{1}{2}
\parbox{1.5cm}{\includegraphics[width=1.5cm]{Figures/qq2qqA.eps}}
-\tfrac{1}{2}
\parbox{1.5cm}{\includegraphics[width=1.5cm]{Figures/qq2qqB.eps}}
-\tfrac{3}{2}
\parbox{1.5cm}{\includegraphics[width=1.5cm]{Figures/qq2qqC.eps}}
-\tfrac{3}{2}
\parbox{1.5cm}{\includegraphics[width=1.5cm]{Figures/qq2qqD.eps}}\ .
\end{gather}
\end{subequations}
At this point we would like to recall once more that the case of universality 
of quark fragmentation as conjectured in Ref.~\cite{Collins:2004nx} is 
included as the situation of vanishing tilde-fragmentation functions.
There would, then, be no gluonic pole scattering cross sections associated 
with quark fragmentation.
Some gluonic pole cross sections associated with quarks are given in the  Tables~\ref{GPCSunpolarized},~\ref{GPCSlongitudinal}
and~\ref{GPCStransverse}.
The others can be obtained from these by using the symmetries described at the end of this section.
For comparison we also give the partonic cross sections, 
Tables~\ref{PCSunpolarized},~\ref{PCSlongitudinal} and~\ref{PCStransverse}
(all cross sections are given for massless quarks).
The gluonic pole cross sections for fermionic scattering have already been calculated in~\cite{Bacchetta:2005rm} and have been reproduced in the tables for completeness.
Using the gluonic pole strengths calculated in the previous section we can now
also give the gluonic pole cross sections for processes with external gluons.
These expressions are important since at high energies gluon scattering 
is expected to be the dominant channel~\cite{Boer:2003tx,Anselmino:2006yq}.

The leading contribution to the parametrization of the gluon correlators~\eqref{GLUONCORR} are transverse to both the momentum of the parent/daughter hadron and its conjugate direction.
The constraint on this conjugate $n$-direction is that it has a nonvanishing overlap with the hadron's momentum,
but can otherwise be chosen arbitrarily.
The expressions for the individual diagrams are not independent of these $n$-vectors.
They enter the expression through the parametrization of the gluon correlators~\eqref{GLUONparametrization} and,
for each external gluon,
there can be a different $n$-vector.
However, in the basic partonic cross sections in Eq.~\eqref{PARTONCROSSSECTION} 
as well as in the gluonic pole cross sections in Eqs.~\eqref{GPCS} all 
the $n$-dependencies cancel. 

To illustrate the role of gluons we take quark-gluon scattering as an example.
If in unpolarized scattering the incoming quark contributes the gluonic pole,
then the corresponding expression in the hadronic scattering cross section will involve the gluonic pole cross section
$d\hat\sigma_{[q]g\rightarrow qg}/d\hat t$ given in Table~\ref{GPCSunpolarized}.
It is obtained by summing the expressions for the diagrams in the first column of Table~\ref{Tqg2qg} after multiplying them with the factors $C_G(q_i)$ in the second column.
In this summation, as with all other gluonic pole cross sections,
all $n$-dependencies cancel.
The situation becomes more complicated when it is the incoming gluon that contributes the gluonic pole.
From equation~\eqref{SINTERKLAAS} and using the results in the third and fourth columns of Table~\ref{Tqg2qg} we see that the gluon-Sivers contribution to the cross section has the form
\begin{equation}
G_T^{(f)}{}^{(1)}(x_1)f_1(x_2)\,
\frac{d\hat\sigma_{[g]q\rightarrow gq}^{(f)}}{d\hat t}\,
\widehat G(z_1)D_1(z_2)
+G_T^{(d)}{}^{(1)}(x_1)f_1(x_2)\,
\frac{d\hat\sigma_{[g]q\rightarrow gq}^{(d)}}{d\hat t}\,
\widehat G(z_1)D_1(z_2)\ .
\end{equation}
Hence, the two possible $T$-odd gluon distribution functions 
(the $x$-even and $x$-odd combinations) 
multiply two distinct gluonic pole cross sections,
one containing the weight factors $C_G^{(f)}$ and one containing the weights $C_G^{(d)}$.
All the gluonic pole cross sections associated with a gluon are given in the tables~\ref{gGPCSunpolarized}-\ref{gGPCStransverse}.

In Table~\ref{Tgg2gg} only the diagrams without 4-gluon vertices have been given.
When 4-gluon vertices are involved, things become slightly more subtle.
At the heart of the problem is that the color structure of the 4-gluon vertex does not factorize from the kinematical part.
This is, then, also the case for the diagram containing the 4-gluon vertex.
This observation has important consequences for the summation of all collinear gluon interactions between the hard and soft parts,
which leads to the gauge-links in the parton correlators.
As an example we consider the diagram
\begin{equation}\label{FOURGLUON}
\parbox{1.5cm}{
\includegraphics[width=1.5cm]{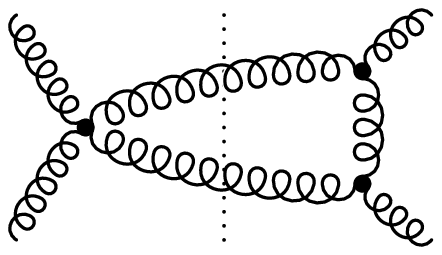}}
=K_1\times\left\{\,\parbox{1.5cm}{
\includegraphics[width=1.5cm]{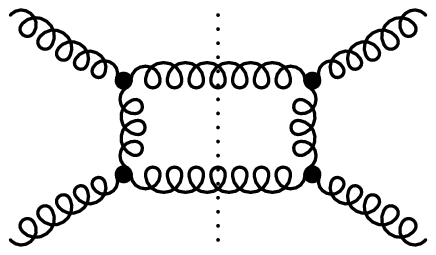}}\,\right\}_c
+K_2\times\left\{\,\parbox{1.5cm}{
\includegraphics[width=1.5cm]{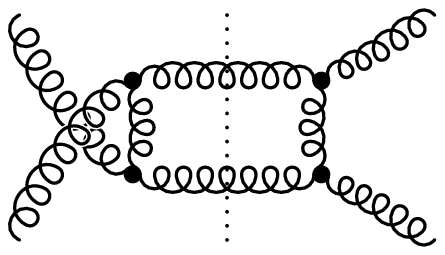}}\,\right\}_c
+K_3\times\left\{\,\parbox{1.5cm}{
\includegraphics[width=1.5cm]{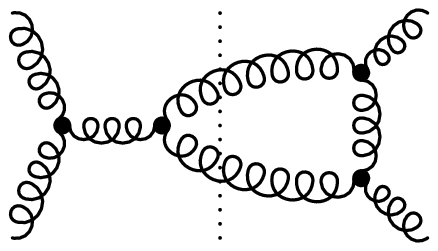}}\,\right\}_c\ .
\end{equation}
On the r.h.s.\ we have made a color-decomposition by inserting the expression of the 4-gluon vertex.
The $K_i$ contain all the kinematical parts and the bracketed diagrams $\{D\}_c$ refer to the \emph{color structure} of that diagram.
Ultimately, it is the color structure that fixes the gauge-link~\cite{Bomhof:2006dp}.
Hence, the three terms on the r.h.s.\ of~\eqref{FOURGLUON} are convoluted with the gauge-invariant correlators of the corresponding color-diagrams
(these are given in Table~8 of Ref.~\cite{Bomhof:2006dp}).
Accordingly, when weighing, the three terms get multiplied by the gluonic pole strengths of the corresponding color-diagrams.
That is, the first term on the r.h.s.\ of~\eqref{FOURGLUON} gets multiplied by the gluonic pole factors in the second row of Table~\ref{Tgg2gg},
the second term gets multiplied by the factors in the last row and the third term gets multiplied by the factors in the fifth row.
As a result, the expression of our diagram as it appears in the calculation of the gluonic pole cross section $d\hat\sigma_{[g]g\rightarrow gg}^{(f)}$ is
\begin{equation}
\parbox{1.5cm}{
\includegraphics[width=1.5cm]{Figures/gg2ggH.eps}}
\longrightarrow\tfrac{1}{2}\times K_1\times\left\{\,\parbox{1.5cm}{
\includegraphics[width=1.5cm]{Figures/gg2ggA.eps}}\,\right\}_c
+1\times K_2\times\left\{\,\parbox{1.5cm}{
\includegraphics[width=1.5cm]{Figures/gg2ggN.eps}}\,\right\}_c
+0\times K_3\times\left\{\,\parbox{1.5cm}{
\includegraphics[width=1.5cm]{Figures/gg2ggE.eps}}\,\right\}_c\ .
\end{equation}
Therefore, the summation in~\eqref{GPCS} is seen to actually run over the \emph{color-decomposed diagrams}.
The expressions for the other diagrams with 4-gluon vertices can be calculated in much the same way and the results are summarized in the tables~\ref{gGPCSunpolarized}-\ref{gGPCStransverse}.

We will end this section by making some observations about the properties of the gluonic pole cross sections.
First of all, the different gluonic pole cross sections cannot be related to one another using the crossing symmetries that exist among the partonic cross sections.
A simple example is $ud{\rightarrow}ud$ versus $u\bar u{\rightarrow}d\bar d$ scattering.
The first process is described by a $t$-channel diagram and the second by an $s$-channel diagram.
Correspondingly, the partonic cross sections can be related by replacing the Mandelstam variables $\hat s$ and $\hat t$ 
(cf.\ Table~\ref{PCSunpolarized}).
Going to the gluonic pole cross sections these two diagrams are multiplied by different gluonic pole factors.
Hence, the gluonic pole cross sections are not related by an 
$\hat s{\leftrightarrow}\hat t$ crossing
(even simpler are the processes $ud{\rightarrow}ud$ and 
$u\bar d{\rightarrow}u\bar d$ that have the same partonic cross sections, 
but different gluonic pole cross sections).
The only crossing symmetry that survives in the gluonic pole cross sections is the substitution $\hat t{\leftrightarrow}\hat u$ when interchanging two partons in the initial or final state.
For a particular partonic process, there are some relations between the gluonic pole cross sections with the gluonic pole associated with different partons.
Using the relations between gluonic pole factors described in the tables in the previous section, 
it is seen that
\begin{subequations}\label{SYMMETRIES}
\begin{alignat}{3}
\frac{d\hat\sigma_{[q_1]q_2\rightarrow ab}}{d\hat t}
&=\frac{d\hat\sigma_{q_1[q_2]\rightarrow ab}}{d\hat t}\ ,&\qquad
\frac{d\hat\sigma_{ab\rightarrow[q_1]q_2}}{d\hat t}
&=\frac{d\hat\sigma_{ab\rightarrow q_1[q_2]}}{d\hat t}\ ,&\qquad
\frac{d\hat\sigma_{[q_1]a\rightarrow q_2b}}{d\hat t}
&=-\frac{d\hat\sigma_{q_1a\rightarrow[q_2]b}}{d\hat t}\ ,\displaybreak[0]\\
\intertext{where $q_1$ and $q_2$ can be any quark or antiquark.
For the gluonic pole cross sections where the gluonic pole is associated with a gluon we have the following relations}
\frac{d\hat\sigma_{[g]g\rightarrow ab}^{(d)}}{d\hat t}
&=-\frac{d\hat\sigma_{g[g]\rightarrow ab}^{(d)}}{d\hat t}\ ,&
\frac{d\hat\sigma_{ab\rightarrow[g]g}^{(d)}}{d\hat t}
&=-\frac{d\hat\sigma_{ab\rightarrow g[g]}^{(d)}}{d\hat t}\ ,&
\frac{d\hat\sigma_{[g]a\rightarrow gb}^{(d)}}{d\hat t}
&=\phantom{-}\frac{d\hat\sigma_{ga\rightarrow[g]b}^{(d)}}{d\hat t}\ ,\\
\frac{d\hat\sigma_{[g]g\rightarrow ab}^{(f)}}{d\hat t}
&=\phantom{-}\frac{d\hat\sigma_{g[g]\rightarrow ab}^{(f)}}{d\hat t}\ ,&
\frac{d\hat\sigma_{ab\rightarrow[g]g}^{(f)}}{d\hat t}
&=\phantom{-}\frac{d\hat\sigma_{ab\rightarrow g[g]}^{(f)}}{d\hat t}\ ,&
\frac{d\hat\sigma_{[g]a\rightarrow gb}^{(f)}}{d\hat t}
&=-\frac{d\hat\sigma_{ga\rightarrow[g]b}^{(f)}}{d\hat t}\ .
\end{alignat}
\end{subequations}
In~\eqref{SYMMETRIES} the $a$ and $b$ can be any colored parton 
(quark, antiquark or gluon).
These symmetries also hold for the polarized cross sections.

Just as for the gluonic pole factors in Eq.~\eqref{QSfactors} there is a relation~\cite{Vogelsang} between the gluonic pole scattering cross sections associated with quarks as calculated in this section and the hard parts $H^{I/F}$ calculated in~\cite{Qiu:1998ia,Kouvaris:2006zy}:
\begin{equation}
\frac{d\hat\sigma_{[q]b\rightarrow cd}}{d\hat t}
=H^I_{qb\rightarrow c}+H^F_{qb\rightarrow c}+H^{F'}_{qb\rightarrow c}\ .
\end{equation}
The $H^{F'}$ for the process $ab{\rightarrow}c_\pi d$ is the same as the $H^F$ of the process $ab{\rightarrow}cd_\pi$ which,
in turn,
can be related to the $H^F$ of the process $ab{\rightarrow}d_\pi c$  by a 
$\hat t{\leftrightarrow}\hat u$ interchange.

\section{Gluonic pole cross sections in $p^\uparrow p{\rightarrow}\pi\pi X$.
\label{example}}

\begin{figure}
\includegraphics{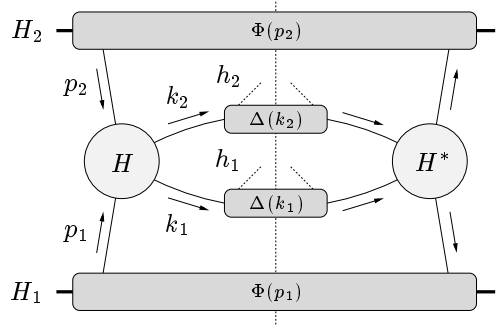}
\caption{The leading order contribution to the cross section of $H_1{+}H_2\rightarrow h_1{+}h_2{+}X$.\label{PPpipi}}
\end{figure}

As an illustration of the appearance of gluonic pole cross sections in 
physical processes we consider inclusive back-to-back pion production 
in proton-proton scattering $p^\uparrow p{\rightarrow}\pi\pi X$,
Figure~\ref{PPpipi}.
In this process, they appear as leading contributions in the azimuthal 
asymmetry considered below.
Using the decompositions~\eqref{DECOMPS} 
$p^\uparrow p{\rightarrow}\pi\pi X$ is given in terms of parton correlators by
\begin{subequations}
\begin{equation}
\langle d\sigma_{h_1h_2}\rangle
=\int d\phi_2\,\frac{d\sigma_{h_1h_2}}{d\phi_2}
=\frac{dx_{1\perp}dx_{2\perp}d\eta_1d\eta_2}{32\pi s}
\frac{d\phi_1}{2\pi}\int\frac{dx_\perp}{x_\perp}\ 
\Sigma(x_1,x_2,z_1,z_2,y)\ ,
\end{equation}
where the momentum fractions are given in terms of the kinematical variables as in Ref.~\cite{Bacchetta:2005rm}.
In that reference it was found that the azimuthal asymmetry
\begin{equation}
\langle\tfrac{1}{2}\sin(\delta\phi)\,d\sigma_{h_1h_2}\rangle
=\int d\phi_2\,\tfrac{1}{2}\sin(\delta\phi)\,\frac{d\sigma_{h_1h_2}}{d\phi_2}
=\frac{dx_{1\perp}dx_{2\perp}d\eta_1d\eta_2}{32\pi s\sqrt s}
\frac{d\phi_1}{2\pi}\int\frac{dx_\perp}{x_\perp^2}\ 
e_{1N}{\cdot}\Sigma_\partial(x_1,x_2,z_1,z_2,y)\ ,\label{WEIGHTED}
\end{equation}
\end{subequations}
contains the leading twist $T$-odd effects.
In these expressions we have used the quantities
\begin{align}
\Sigma(x_1,x_2,z_1,z_2,y)
&=\sum\tr\big\{\,\Phi(x_1)\Phi(x_2)\Delta(z_1)\Delta(z_2)\,
H\,H^*\,\big\}\ ,\\
\Sigma_\partial^\alpha(x_1,x_2,z_1,z_2,y)
&=\sum\tr\big\{\,\big[\Phi_\partial^\alpha(x_1)\Phi(x_2)\Delta(z_1)\Delta(z_2)
{+}\Phi(x_1)\Phi_\partial^\alpha(x_2)\Delta(z_1)\Delta(z_2)\nonumber\\
&\mspace{100mu}
{-}\Phi(x_1)\Phi(x_2)\Delta_\partial^\alpha(z_1)\Delta(z_2)
{-}\Phi(x_1)\Phi(x_2)\Delta(z_1)\Delta_\partial^\alpha(z_2)\big]\,
H\,H^*\,\big\}\\
&\equiv\Sigma_{1\,\partial}^{\alpha}+\Sigma_{2\,\partial}^{\alpha}
-\Sigma_{1^\prime\,\partial}^{\alpha}-\Sigma_{2^\prime\,\partial}^{\alpha}\ .
\end{align}
The quantity $H{=}H(x_1,x_2,z_1,z_2)$ is the collinear hard part
and the summation runs over all parton correlators as given in 
appendix~\ref{CORRS}.

For the parametrizations of the different correlators we use the conventions 
of Ref.~\cite{Bacchetta:2005rm} and Eqs.~\eqref{GLUONparametrization}
and~\eqref{GPMelement}.
Inserting these parametrizations in the expressions for the cross sections, 
one obtains
\begin{subequations}\label{UNPOLARIZED}
\begin{gather}
\frac{\Sigma(x_1,x_2,z_1,z_2,y)}{32\pi s}\nonumber\\
\mspace{50mu}
=\sum_{\begin{subarray}{c}\text{(anti)quark}\\ \text{flavors}\end{subarray}}\Big\{\,
f_1(x_1)f_1(x_2)\,\frac{\hat s}{2}
\frac{d\hat \sigma_{q_1q_2\rightarrow q_3q_4}}{d\hat t}\,
D_1(z_1)D_1(z_2)\\[-4mm]
\mspace{190mu}
+f_1(x_1)G(x_2)\,\frac{\hat s}{2}
\frac{d\hat \sigma_{qg\rightarrow qg}}{d\hat t}\,
D_1(z_1)\widehat G(z_2)\ 
+\big(K_1{\leftrightarrow}K_2\big)\displaybreak[0]\\
\mspace{190mu}
+G(x_1)f_1(x_2)\,\frac{\hat s}{2}
\frac{d\hat \sigma_{gq\rightarrow gq}}{d\hat t}\,
\widehat G(z_1)D_1(z_2)\ 
+\big(K_1{\leftrightarrow}K_2\big)\displaybreak[0]\\
\mspace{190mu}
+f_1(x_1)\bar f_1(x_2)\,\frac{\hat s}{2}
\frac{d\hat \sigma_{q\bar q\rightarrow gg}}{d\hat t}\,
\widehat G(z_1)\widehat G(z_2)\displaybreak[0]\\
\mspace{190mu}
+G(x_1)G(x_2)\,\frac{\hat s}{2}
\frac{d\hat \sigma_{gg\rightarrow q\bar q}}{d\hat t}\,
D_1(z_1)\bar D_1(z_2)\\
\mspace{190mu}
+G(x_1)G(x_2)\,\frac{\hat s}{2}
\frac{d\hat \sigma_{gg\rightarrow gg}}{d\hat t}\,
\widehat G(z_1)\widehat G(z_2)\ \Big\}\ ,
\end{gather}
\end{subequations}
where the summation runs over all quarks and antiquarks
(we have suppressed the parton indices on the distribution and fragmentation functions).
For the polarized cross section we get
(the quark-scattering contributions were already obtained in Eqs.~(30)-(32) and (D1)-(D3) of Ref.~\cite{Bacchetta:2005rm}, 
but with a wrong overall sign)
\begin{subequations}\label{POLARIZED}
\begin{gather}
-\frac{e_{1N}\cdot\Sigma_{1\,\partial}(x_1,x_2,z_1,z_2,y)}
{32\pi s\,M_1\cos(\phi_1{-}\phi_S)}\nonumber\\
\mspace{50mu}
=\sum_{\begin{subarray}{c}\text{(anti)quark}\\ \text{flavors}\end{subarray}}\bigg\{\ 
f_{1T}^{\perp(1)}(x_1)f_1(x_2)\,\frac{\hat s}{2}
\frac{d\hat\sigma_{[q_1]q_2\rightarrow q_3q_4}}{d\hat t}\,
D_1(z_1)D_1(z_2)\label{QuarkSiversA}\\[-3mm]
\mspace{190mu}
+f_{1T}^{\perp(1)}(x_1)G(x_2)\,\frac{\hat s}{2}
\frac{d\hat\sigma_{[q]g\rightarrow qg}}{d\hat t}\,
D_1(z_1)\widehat G(z_2)\ 
+\big(K_1{\leftrightarrow}K_2\big)\label{QuarkSiversB}\displaybreak[0]\\
\mspace{190mu}
+f_{1T}^{\perp(1)}(x_1)\bar f_1(x_2)\,\frac{\hat s}{2}
\frac{d\hat\sigma_{[q]\bar q\rightarrow gg}}{d\hat t}\,
\widehat G(z_1)\widehat G(z_2)\label{QuarkSiversC}\displaybreak[0]\\
\mspace{190mu}
+G_T^{(d)}{}^{(1)}(x_1)f_1(x_2)\,
\frac{\hat s}{2}\frac{d\hat\sigma_{[g]q\rightarrow gq}^{(d)}}{d\hat t}\,
\widehat G(z_1)D_1(z_2)\ 
+\big(K_1{\leftrightarrow}K_2\big)\label{GluonSiversA}\displaybreak[0]\\
\mspace{190mu}
+G_T^{(f)}{}^{(1)}(x_1)f_1(x_2)\,
\frac{\hat s}{2}\frac{d\hat\sigma_{[g]q\rightarrow gq}^{(f)}}{d\hat t}\,
\widehat G(z_1)D_1(z_2)\ 
+\big(K_1{\leftrightarrow}K_2\big)\label{GluonSiversB}\displaybreak[0]\\
\mspace{190mu}
+G_T^{(d)}{}^{(1)}(x_1)G(x_2)\,
\frac{\hat s}{2}\frac{d\hat\sigma_{[g]g\rightarrow q\bar q}^{(d)}}{d\hat t}\,
D_1(z_1)\bar D_1(z_2)\label{GluonSiversC}\displaybreak[0]\\
\mspace{190mu}
+G_T^{(f)}{}^{(1)}(x_1)G(x_2)\,
\frac{\hat s}{2}\frac{d\hat\sigma_{[g]g\rightarrow q\bar q}^{(f)}}{d\hat t}\,
D_1(z_1)\bar D_1(z_2)\label{GluonSiversD}\displaybreak[0]\\
\mspace{190mu}
+G_T^{(d)}{}^{(1)}(x_1)G(x_2)\,
\frac{\hat s}{2}\frac{d\hat\sigma_{[g]g\rightarrow gg}^{(d)}}{d\hat t}\,
\widehat G(z_1)\widehat G(z_2)\label{GluonSiversE}\\
\mspace{190mu}
+G_T^{(f)}{}^{(1)}(x_1)G(x_2)\,
\frac{\hat s}{2}\frac{d\hat\sigma_{[g]g\rightarrow gg}^{(f)}}{d\hat t}\,
\widehat G(z_1)\widehat G(z_2)\ \Big\}\ ,\label{GluonSiversF}
\end{gather}
\end{subequations}
\begin{subequations}
\begin{gather}
-\frac{e_{1N}\cdot\Sigma_{2\,\partial}(x_1,x_2,z_1,z_2,y)}
{32\pi s\,M_2\cos(\phi_1{-}\phi_S)}\nonumber\\
\mspace{50mu}
=\sum_{\begin{subarray}{c}\text{(anti)quark}\\ \text{flavors}\end{subarray}}\bigg\{\ 
h_1(x_1)h_1^{\perp(1)}(x_2)\,\frac{\hat s}{2}
\frac{d\Delta\hat\sigma_{q_1^\uparrow[q_2^\uparrow]
\rightarrow q_3q_4}}{d\hat t}\,
D_1(z_1)D_1(z_2)\label{BoerMuldersA}\displaybreak[0]\\[-3mm]
\mspace{190mu}
+h_1(x_1)\bar h{}_1^{\perp(1)}(x_2)\,\frac{\hat s}{2}
\frac{d\Delta\hat\sigma_{q^\uparrow[\bar q{}^\uparrow]\rightarrow gg}}
{d\hat t}\,\widehat G(z_1)\widehat G(z_2)\ \Big\}\ ,\label{BoerMuldersB}
\end{gather}
\end{subequations}
\begin{subequations}\label{CollEffect}
\begin{gather}
-\frac{e_{1N}\cdot\Sigma_{1^\prime\,\partial}(x_1,x_2,z_1,z_2,y)}
{32\pi s\,M_{h_1}\cos(\phi_1{-}\phi_S)}\nonumber\\
\mspace{50mu}
=\sum_{\begin{subarray}{c}\text{(anti)quark}\\ \text{flavors}\end{subarray}}\bigg\{\ 
h_1(x_1)f_1(x_2)\,\frac{\hat s}{2}
\frac{d\Delta\hat\sigma_{q_1^\uparrow q_2^{\phantom{\uparrow}}\rightarrow 
q_3^\uparrow q_4^{\phantom{\uparrow}}}}{d\hat t}\,
H_1^{\perp(1)}(z_1)D_1(z_2)
\label{CollinsA}\displaybreak[0]\\[-3mm]
\mspace{190mu}
+h_1(x_1)f_1(x_2)\,\frac{\hat s}{2}
\frac{d\Delta\hat\sigma_{q_1^\uparrow q_2^{\phantom{\uparrow}}
\rightarrow[q_3^\uparrow]q_4^{\phantom{\uparrow}}}}{d\hat t}\,
\widetilde H{}_1^{\perp(1)}(z_1)D_1(z_2)
\label{CollinsB}\displaybreak[0]\\
\mspace{190mu}
+h_1(x_1)G(x_2)\,\frac{\hat s}{2}
\frac{d\Delta\hat\sigma_{q^\uparrow g\rightarrow q^\uparrow g}}{d\hat t}\,
H_1^{\perp(1)}(z_1)\widehat G(z_2)
\label{CollinsC}\\
\mspace{190mu}
+h_1(x_1)G(x_2)\,\frac{\hat s}{2}
\frac{d\Delta\hat\sigma_{q^\uparrow g\rightarrow[q^\uparrow]g}}{d\hat t}\,
\widetilde H_1^{\perp(1)}(z_1)\widehat G(z_2)\ \Big\}\ .\label{CollinsD}
\end{gather}
\end{subequations}
With $K_1{\leftrightarrow}K_2$ we denote an interchange of the two outgoing hadrons.
For this process 
($H_1^\uparrow H_2^{\phantom{\uparrow}}{\rightarrow}h_1h_2X)$ 
the contraction $e_{1N}{\cdot}\Sigma_{2^\prime\,\partial}$ is obtained from
$e_{1N}{\cdot}\Sigma_{1^\prime\,\partial}$ by such an interchange.
The gluonic pole scattering cross sections appearing in these expressions 
can be read from the tables in the previous section,
either directly or indirectly by using the symmetry relations described 
at the end of that section.
In Eqs.~\eqref{UNPOLARIZED}-\eqref{CollEffect} the distribution and 
fragmentation functions appear convoluted, 
as the momentum fractions are functions of the scaled parton perpendicular momentum
$x_\perp{\equiv}2|k_{2\perp}|/\sqrt s$ 
(see section IV of Ref.~\cite{Bacchetta:2005rm}).
The weighed cross section for back-to-back jet production
$p^\uparrow p{\rightarrow}j_1j_2X$ can simply be obtained from the 
expression given above by taking $D_1(z_i){=}\widehat G(z_i){=}\delta(z_i{-}1)$
and by setting all other fragmentation functions to zero.
With these substitutions the distribution functions are no longer 
convoluted~\cite{Bacchetta:2005rm}.
In Fig.~\ref{RelDifferences} some of the gluonic pole cross sections 
appearing in the expressions above are compared to the partonic cross sections.
It should be noted that in the azimuthal asymmetry in Eq.~\eqref{WEIGHTED} 
each of the gluonic pole cross sections is weighed with a different 
combination of distribution and fragmentation functions and they are added and integrated over phase-space.
In this paper we have not given quantitative results for the hadronic cross sections.
This requires input of distribution and fragmentation functions for
which one has to use existing parameterizations for the know functions~\cite{Gluck:1998xa,Leader:2005ci,Hirai:2006sr,Thorne:2006zu}
and model input~\cite{Signal:1989yc,Barone:1993iq,Jakob:1997wg,Diakonov:1998ze,Goeke:2000wv,Gluck:2000dy,Bacchetta:2001di,Wakamatsu:2000fd,Bacchetta:2002tk,Yang:2002gh,Efremov:2004ph} for the unknown ones. 
Hence, it remains to be seen whether the use of gluonic pole cross sections instead of partonic cross sections leads to observable differences.
We will leave that as the subject of future publications.

\section{Conclusions}

We have shown how single spin asymmetries at tree-level can be described as convolutions of parton distribution and fragmentation functions and a hard cross section in a way similar as is usual for spin-averaged cross sections. 
The perturbatively calculable hard cross sections, 
referred to as {\em gluonic pole cross sections}, 
contain the standard squared amplitudes in the diagrammatic expansion of the process multiplied with calculable {\em gluonic pole factors}. 
These depend only on the color flow in the Feynman diagrams.

For the single spin asymmetries in SIDIS and the Drell-Yan
process, the sign switch of the Sivers effects is a direct consequence 
of the appearance of future (SIDIS) and past (Drell-Yan) pointing 
gauge-links in the transverse momentum dependent distribution functions. 
These signs are examples of gluonic pole factors. 
In this case only one squared amplitude contributes and, consequently, 
the gluonic pole cross sections and the standard partonic cross section are simply proportional. 

Going beyond the simplest electroweak processes such as SIDIS, DY or $e^+e^-$ annihilation,
the plus/minus signs in the Sivers effect generalize to arbitrary, 
but calculable factors.
These are the gluonic pole factors that must be included when one encounters $p_T$-weighted correlators (the transverse moments),
e.g.\ appearing in single spin azimuthal asymmetries in hadronic processes. 
The gluonic pole factors are completely fixed by the color structure 
of the hard partonic 
subprocess. 
For that reason they are naturally associated with it.
When there are several diagrams that could contribute to the asymmetry, 
each diagram can have a different gluonic pole factor. 
Hence one in general does not get the simple proportionality 
of gluonic pole and standard cross sections. 
Instead one has a weighted sum of squared amplitudes in the gluonic pole cross section rather than the ordinary sum in the standard cross sections.
These weighted sums are properly color gauge-invariant.
In the case that the $T$-odd effect is associated with a gluon there is a 
doubling of the $T$-odd gluon distribution (or fragmentation) functions that appear in the parametrizations of the gluonic pole matrix elements.
These correspond to the two different ways of constructing color-singlet gluonic pole matrix elements.

In this paper we have calculated all gluonic pole factors and gluonic pole scattering cross sections
that can contribute to single spin asymmetries in hadron-hadron scattering.
These involve two-to-two hard subprocesses of colored particles.
We have explicitly worked out the case of inclusive back-to-back pion and jet
production 
($p^\uparrow p{\rightarrow}\pi\pi X$, $p^\uparrow p{\rightarrow}j_1j_2X$)
relevant for ongoing RHIC experiments.
In such processes it is possible to define azimuthal asymmetries in which the leading contribution contains gluonic pole cross sections. 
As a caveat we mention that in this paper we have assumed that the hadronic cross sections of these processes at leading twist factorize in soft correlators and hard partonic amplitudes.
This is an assumption that remains to be proved.
However, we have confidence in the procedure,
since they are two-scale processes~\cite{Boer:2003tx,Bacchetta:2005rm} and because our conclusions hold for the tree-level contribution where any soft-factors will (probably) be unity.
The gluonic pole cross sections may also play a role in one-pion inclusive processes 
($p^\uparrow p{\rightarrow}\pi X$), but then certainly in combination
with subleading twist $p_T$-integrated 
functions~\cite{Anselmino:1994tv,Anselmino:1998yz}.

\begin{acknowledgments}
We acknowledge discussions with A.~Bacchetta, D.~Boer and W.~Vogelsang. 
This work is included in the research program of the EU Integrated
Infrastructure Initiative Hadron Physics (RII3-CT-2004-506078) and is in
part supported by the Foundation for Fundamental 
Research of Matter (FOM) and the National Organization for Scientific 
Research (NWO).
\end{acknowledgments}

\appendix

\section{correlators\label{CORRS}}

Using the decompositions given in~\eqref{DECOMPS} all transverse momentum dependent distribution and fragmentation functions can be found as specific projections of the correlators
defined on the light-front LF ($\xi{\cdot}n{\equiv}0$) \cite{Soper:1976jc,Soper:1979fq,Ralston:1979ys,Collins:1981uw,
Collins:1981tt,Collins:1982wa,Jaffe:1991ra,Mulders:1995dh}
\begin{subequations}\label{QUARKCORR}
\begin{gather}
\Phi^{[\mathcal U]}(x,p_T)
=\int\frac{d(\xi{\cdot}P)d^2\xi_T}{(2\pi)^3}\ e^{ip{\cdot}\xi}\,
\langle P{,}S|\,\overline\psi(0)\,\mathcal U_{[0;\xi]}\,
\psi(\xi)\,|P{,}S\rangle\,\big\rfloor_{\text{LF}}\ ,\\
\Delta^{[\mathcal U]}\big(\tfrac{1}{z},k_T\big)
=\sum_X\frac{1}{4z}\int\frac{d(\xi{\cdot}K)d^2\xi_T}{(2\pi)^3}\ e^{-ik{\cdot}\xi}\,
\langle 0|\,\mathcal U_0\,\psi(0)\,|h{,}X\rangle
\langle h{,}X|\,\overline\psi(\xi)\,
\mathcal U_\xi\,|0\rangle\,\big\rfloor_{\text{LF}}\ ,\label{QUARKCORRb}
\end{gather}
\end{subequations}
in the case of quarks,
\begin{subequations}
\begin{gather}
\bar\Phi{}^{[\mathcal U]}(x,p_T)
=\int\frac{d(\xi{\cdot}P)d^2\xi_T}{(2\pi)^3}\ e^{ip{\cdot}\xi}\,
\langle P{,}S|\,\psi(0)\,\overline\psi(\xi)\,
\mathcal U_{[\xi;0]}\,|P{,}S\rangle\,\big\rfloor_{\text{LF}}\ ,\\
\bar\Delta{}^{[\mathcal U]}\big(\tfrac{1}{z},k_T\big)
=\sum_X\frac{1}{4z}\int\frac{d(\xi{\cdot}K)d^2\xi_T}{(2\pi)^3}\ e^{-ik{\cdot}\xi}\,
\langle 0|\,\overline\psi(0)\,\mathcal U_0\,|h{,}X\rangle
\langle h{,}X|\,\mathcal U_\xi\psi(\xi)\,\,|0\rangle\,\big\rfloor_{\text{LF}}\ ,
\end{gather}
\end{subequations}
in the case of antiquarks and
\begin{subequations}\label{GLUONCORR}
\begin{gather}
\Gamma^{[\mathcal U\mathcal U']\mu\nu}(x,p_T)
=\frac{n_\rho n_\sigma}{(p{\cdot}n)^2}
\int\frac{d(\xi{\cdot}P)d^2\xi_T}{(2\pi)^3}\ e^{ip{\cdot}\xi}\,
\langle P{,}S|\,F^{\mu\rho}(0)\,\mathcal U_{[0;\xi]}\,
F^{\nu\sigma}(\xi)\,\mathcal U_{[\xi;0]}'\,
|P{,}S\rangle\,\big\rfloor_{\text{LF}}\ ,\label{GLUONCORRa}\\
\widehat\Gamma{}^{[\mathcal U\mathcal U']\mu\nu}\big(\tfrac{1}{z},k_T\big)
=\frac{n_{h\rho}n_{h\sigma}}{(k{\cdot}n_h)^2}
\sum_X\frac{1}{4z}\int\frac{d(\xi{\cdot}K)d^2\xi_T}{(2\pi)^3}\
e^{-ik{\cdot}\xi}\,
\langle 0|\,\mathcal U_0'\,F^{\mu\rho}(0)\,\mathcal U_0\,|h{,}X\rangle
\langle h{,}X|\,\mathcal U_\xi\,F^{\nu\sigma}(\xi)\,
\mathcal U_\xi'\,|0\rangle\,\big\rfloor_{\text{LF}}\ ,
\end{gather}
\end{subequations}
in the case of gluons.
All correlators are color traced (suppressed in the expressions above).
In the fragmentation correlators the $\mathcal U_0$ and $\mathcal U_\xi$ 
refer to those parts of the gauge-link $\mathcal U$ that connect to the 
space-time points $0$ and $\xi$, respectively.
Collinear distribution and fragmentation functions are projections of 
the collinear correlators 
\begin{subequations}
\begin{alignat}{2}
\Phi(x)
&=\int d^2p_T\ \Phi^{[\mathcal U]}(x,p_T)\ ,&\qquad\qquad
\Delta\big(\tfrac{1}{z}\big)
&=z^2\int d^2k_T\ \Delta^{[\mathcal U]}\big(\tfrac{1}{z},k_T\big)\ ,
\\
\Phi_\partial^\alpha(x)
&=\int d^2p_T\ p_T^\alpha\ \Phi^{[\mathcal U]}(x,p_T)\ ,&
\Delta_\partial^\alpha\big(\tfrac{1}{z}\big)
&=z^2\int d^2k_T\ k_T^\alpha\ \Delta^{[\mathcal U]}\big(\tfrac{1}{z},k_T\big)\ ,
\label{kaassoufle-app}
\end{alignat}
\end{subequations}
and similarly for the antiquark and gluon correlators.
Other useful correlators defined on the lightcone LC 
($\xi{\cdot}n{\equiv}\xi_T{\equiv}0$) are those with additional covariant derivatives
\begin{subequations}
\begin{gather}
\Phi_D^\alpha(x)
=\int\frac{d(\xi{\cdot}P)}{2\pi}\ e^{ix(\xi\cdot P)}\,
\langle P{,}S|\,\overline\psi(0)\,U_{[0;\xi]}^n\,iD^\alpha(\xi)\,
\psi(\xi)\,|P{,}S\rangle\,\big\rfloor_{\text{LC}}\ ,\\
\Delta_D^\alpha\big(\tfrac{1}{z}\big)
=\sum_X\frac{z}{4}
\int\frac{d(\xi{\cdot}K)}{2\pi}\ e^{-i\frac{1}{z}(\xi{\cdot}K)}\,
\langle0|\,U^{n_h}_{[\zeta;0]}\,\psi(0)\,|h{,}X\rangle
\langle h{,}X|\,\overline{\big(iD^\alpha(\xi)\psi(\xi)\big)}\,
U^{n_h}_{[\xi;\zeta]}\,|0\rangle\,\big\rfloor_{\text{LC}}\ ,\displaybreak[0]\\
\bar\Phi{}_D^\alpha(x)
=-\int\frac{d(\xi{\cdot}P)}{2\pi}\ e^{ix(\xi\cdot P)}\,
\langle P{,}S|\,\psi(0)\,\overline{\big(iD^\alpha(\xi)\psi(\xi)\big)}\,
U_{[\xi;0]}^n\,|P{,}S\rangle\,\big\rfloor_{\text{LC}}\ ,\displaybreak[0]\\
\bar\Delta{}_D^\alpha\big(\tfrac{1}{z}\big)
=-\sum_X\frac{z}{4}
\int\frac{d(\xi{\cdot}K)}{2\pi}\ e^{-i\frac{1}{z}(\xi{\cdot}K)}\,
\langle0|\,\overline\psi(0)\,U^{n_h}_{[0;\zeta]}\,|h{,}X\rangle
\langle h{,}X|\,U^{n_h}_{[\zeta;\xi]}\,iD^\alpha(\xi)\psi(\xi)\,
|0\rangle\,\big\rfloor_{\text{LC}}\ ,\displaybreak[0]\\
\Gamma_D^{\mu\nu;\alpha}(x)
=\frac{n_\rho n_\sigma}{(p{\cdot}n)^2}
\int\frac{d(\xi{\cdot}P)}{2\pi}\ e^{ix(\xi{\cdot}P)}\,
\langle P{,}S|\,F^{\mu\rho}(0)\,U^n_{[0;\xi]}\,
\big[iD^\alpha(\xi),F^{\nu\sigma}(\xi)\big]\,
U^n_{[\xi;0]}\,|P{,}S\rangle\,\big\rfloor_{\text{LC}}\ ,\\
\widehat\Gamma{}_D^{\mu\nu;\alpha}\big(\tfrac{1}{z}\big)
=-\frac{n_{h\rho}n_{h\sigma}}{(k{\cdot}n_h)^2}
\sum_X\frac{z}{4}\int\frac{d(\xi{\cdot}K)}{2\pi}\ 
e^{-i\frac{1}{z}(\xi{\cdot}K)}\nonumber\\
\mspace{200mu}
\times\ 
\langle0|\,U^{n_h}_{[\zeta;0]}\,F^{\mu\rho}(0)\,
U^{n_h}_{[0;\zeta']}\,|h,X\rangle
\langle h{,}X|\,U^{n_h}_{[\zeta';\xi]}\,
\big[iD^\alpha(\xi),F^{\nu\sigma}(\xi)\big]\,
U^{n_h}_{[\xi;\zeta]}\,|0\rangle\,\big\rfloor_{\text{LC}}\ ,
\end{gather}
\end{subequations}
and gluon fields 
(with $F^{n\alpha}{\equiv}F^{\mu\alpha}n_\mu/(P{\cdot}n)$
and $F^{n_h\alpha}{\equiv}F^{\mu\alpha}n_{h\mu}/(K{\cdot}n_h)$)
\begin{subequations}
\begin{gather}
\Phi_G^\alpha(x,x{-}x')
=\int\frac{d(\xi{\cdot}P)}{2\pi}\frac{d(\eta{\cdot}P)}{2\pi}\ 
e^{i(x-x')(\xi\cdot P)}e^{ix'(\eta\cdot P)}\,
\langle P{,}S|\,\overline\psi(0)\,U_{[0;\eta]}^n\,gF^{n\alpha}(\eta)\,
U_{[\eta;\xi]}^n\,\psi(\xi)\,|P{,}S\rangle\,\big\rfloor_{\text{LC}}\ ,\\
\Delta_G^\alpha\big(\tfrac{1}{z},\tfrac{1}{z}{-}\tfrac{1}{z'}\big)
=\sum_X\frac{z}{4}\int\frac{d(\xi{\cdot}K)}{2\pi}\frac{d(\eta{\cdot}K)}{2\pi}\ 
e^{-i(\frac{1}{z}-\frac{1}{z'})(\xi{\cdot}K)}
e^{-i\frac{1}{z'}(\eta{\cdot}K)}\,\nonumber\\
\mspace{160mu}\times\ 
\langle0|\,U^{n_h}_{[\zeta;0]}\,\psi(0)\,|h{,}X\rangle
\langle h{,}X|\,\overline\psi(\xi)\,U^{n_h}_{[\xi;\eta]}\,gF^{n_h\alpha}(\eta)\,
U^{n_h}_{[\eta;\zeta]}\,|0\rangle\,\big\rfloor_{\text{LC}}\ ,\displaybreak[0]\\
\bar\Phi{}_G^\alpha(x,x{-}x')
=-\int\frac{d(\xi{\cdot}P)}{2\pi}\frac{d(\eta{\cdot}P)}{2\pi}\ 
e^{i(x-x')(\xi\cdot P)}e^{ix'(\eta\cdot P)}\,
\langle P{,}S|\,\psi(0)\,\overline\psi(\xi)\,U_{[\xi;\eta]}^n\,
gF^{n\alpha}(\eta)\,U_{[\eta;0]}^n\,
|P{,}S\rangle\,\big\rfloor_{\text{LC}}\ ,\displaybreak[0]\\
\bar\Delta{}_G^\alpha\big(\tfrac{1}{z},\tfrac{1}{z}{-}\tfrac{1}{z'}\big)
=-\sum_X\frac{z}{4}\int\frac{d(\xi{\cdot}K)}{2\pi}\frac{d(\eta{\cdot}K)}{2\pi}\ 
e^{-i(\frac{1}{z}-\frac{1}{z'})(\xi{\cdot}K)}
e^{-i\frac{1}{z'}(\eta{\cdot}K)}\,\nonumber\\
\mspace{160mu}\times\ 
\langle0|\,\overline\psi(0)\,U^{n_h}_{[0;\zeta]}\,|h{,}X\rangle
\langle h{,}X|\,U^{n_h}_{[\zeta;\eta]}\,gF^{n_h\alpha}(\eta)\,
U^{n_h}_{[\eta;\xi]}\,\psi(\xi)\,|0\rangle\,\big\rfloor_{\text{LC}}\ ,
\displaybreak[0]\\
\Gamma_G^{(d)}{}^{\mu\nu;\alpha}(x,x{-}x')
=\frac{n_\rho n_\sigma}{(p{\cdot}n)^2}
\int\frac{d(\xi{\cdot}P)}{2\pi}\frac{d(\eta{\cdot}P)}{2\pi}\ 
e^{i(x{-}x')(\xi{\cdot}P)}e^{ix'(\eta{\cdot}P)}\nonumber\\
\mspace{160mu}\times\ 
\langle P{,}S|\,F^{\mu\rho}(0)\,
\big\{U_{[0,\eta]}^ngF^{n\alpha}(\eta)U_{[\eta,0]}^n,
U_{[0,\xi]}^nF^{\nu\sigma}(\xi)U_{[\xi,0]}^n\big\}\,
|P{,}S\rangle\,\big\rfloor_{\text{LC}}\ ,\displaybreak[0]\\
\Gamma_G^{(f)}{}^{\mu\nu;\alpha}(x,x{-}x')
=\frac{n_\rho n_\sigma}{(p{\cdot}n)^2}
\int\frac{d(\xi{\cdot}P)}{2\pi}\frac{d(\eta{\cdot}P)}{2\pi}\ 
e^{i(x{-}x')(\xi{\cdot}P)}e^{ix'(\eta{\cdot}P)}\nonumber\\
\mspace{160mu}\times\ 
\langle P{,}S|\,F^{\mu\rho}(0)\,
\big[U_{[0,\eta]}^ngF^{n\alpha}(\eta)U_{[\eta,0]}^n,
U_{[0,\xi]}^nF^{\nu\sigma}(\xi)U_{[\xi,0]}^n\big]\,
|P{,}S\rangle\,\big\rfloor_{\text{LC}}\ ,\label{sinaasappelsap}\displaybreak[0]\\
\widehat\Gamma{}_G^{(d)}{}^{\mu\nu;\alpha}
\big(\tfrac{1}{z},\tfrac{1}{z}{-}\tfrac{1}{z'}\big)
=-\frac{n_{h\rho}n_{h\sigma}}{(k{\cdot}n_h)^2}
\sum_X\frac{z}{4}\int\frac{d(\xi{\cdot}K)}{2\pi}\frac{d(\eta{\cdot}K)}{2\pi}\ 
e^{-i(\frac{1}{z}-\frac{1}{z'})(\xi{\cdot}K)}
e^{-i\frac{1}{z'}(\eta{\cdot}K)}\nonumber\\
\mspace{160mu}
\times\ \langle0|\,U^{n_h}_{[\zeta;0]}
F^{\mu\rho}(0)U^{n_h}_{[0;\zeta]}\,|h{,}X\rangle
\langle h{,}X|\,\big\{U^{n_h}_{[\zeta;\eta]}gF^{n_h\alpha}(\eta)
U^{n_h}_{[\eta;\zeta]},U^{n_h}_{[\zeta;\xi]}F^{\nu\sigma}(\xi)
U^{n_h}_{[\xi;\zeta]}\big\}\,|0\rangle\,\big\rfloor_{\text{LC}}\ ,\\
\widehat\Gamma{}_G^{(f)}{}^{\mu\nu;\alpha}
\big(\tfrac{1}{z},\tfrac{1}{z}{-}\tfrac{1}{z'}\big)
=-\frac{n_{h\rho}n_{h\sigma}}{(k{\cdot}n_h)^2}
\sum_X\frac{z}{4}\int\frac{d(\xi{\cdot}K)}{2\pi}\frac{d(\eta{\cdot}K)}{2\pi}\ 
e^{-i(\frac{1}{z}-\frac{1}{z'})(\xi{\cdot}K)}
e^{-i\frac{1}{z'}(\eta{\cdot}K)}\nonumber\\
\mspace{160mu}
\times\ \langle0|\,U^{n_h}_{[\zeta;0]}F^{\mu\rho}(0)
U^{n_h}_{[0;\zeta]}\,|h{,}X\rangle
\langle h{,}X|\,\big[U^{n_h}_{[\zeta;\eta]}gF^{n_h\alpha}(\eta)
U^{n_h}_{[\eta;\zeta]},U^{n_h}_{[\zeta;\xi]}F^{\nu\sigma}(\xi)
U^{n_h}_{[\xi;\zeta]}\big]\,|0\rangle\,\big\rfloor_{\text{LC}}\ .
\end{gather}
\end{subequations}
The weighted parton correlators can be related to the correlators defined above:
\begin{subequations}
\begin{gather}
\Phi_\partial^{[\mathcal U]\alpha}(x)
=\widetilde\Phi{}_\partial^\alpha(x)
+C_G^{[\mathcal U]}\,\pi\Phi_G^\alpha(x,x)\ ,\\
\Delta_\partial^{[\mathcal U]\alpha}\big(\tfrac{1}{z}\big)
=\widetilde\Delta{}_\partial^\alpha\big(\tfrac{1}{z}\big)
+C_G^{[\mathcal U]}\,
\pi\Delta_G^\alpha\big(\tfrac{1}{z},\tfrac{1}{z}\big)\ ,\displaybreak[0]\\
\bar\Phi{}_\partial^{[\mathcal U]\alpha}(x)
=\widetilde{\bar\Phi}{}_\partial^\alpha(x)
+C_G^{[\mathcal U]}\,\pi\bar\Phi{}_G^\alpha(x,x)\ ,\displaybreak[0]\\
\bar\Delta{}_\partial^{[\mathcal U]\alpha}\big(\tfrac{1}{z}\big)
=\widetilde{\bar\Delta}{}_\partial^\alpha\big(\tfrac{1}{z}\big)
+C_G^{[\mathcal U]}\,\pi\bar\Delta{}_G^\alpha
\big(\tfrac{1}{z},\tfrac{1}{z}\big)\ ,\displaybreak[0]\\
\Gamma_\partial^{[\mathcal U\mathcal U']\alpha}(x)
=\widetilde\Gamma{}_\partial^\alpha(x)
+C_G^{(f)}{}^{[\mathcal U\mathcal U']}\,\pi\Gamma_G^{(f)}{}^\alpha(x,x)
+C_G^{(d)}{}^{[\mathcal U\mathcal U']}\,\pi\Gamma_G^{(d)}{}^\alpha(x,x)\ ,\\
\widehat\Gamma{}_\partial^{[\mathcal U\mathcal U']\alpha}\big(\tfrac{1}{z}\big)
=\widetilde{\widehat\Gamma}{}_\partial^\alpha\big(\tfrac{1}{z}\big)
+C_G^{(f)}{}^{[\mathcal U\mathcal U']}\,
\pi\widehat\Gamma_G^{(f)}{}^\alpha\big(\tfrac{1}{z},\tfrac{1}{z}\big)
+C_G^{(d)}{}^{[\mathcal U\mathcal U']}\,
\pi\widehat\Gamma_G^{(d)}{}^\alpha\big(\tfrac{1}{z},\tfrac{1}{z}\big)\ ,
\end{gather}
\end{subequations}
with
\begin{subequations}
\begin{gather}
\widetilde\Phi{}_\partial^\alpha(x)
=\Phi_D^\alpha(x)
-\int dx'\ P\frac{i}{x'}\ \Phi_G^\alpha(x,x{-}x')\ ,\\
\widetilde\Delta{}_\partial^\alpha\big(\tfrac{1}{z}\big)
=\Delta_D^\alpha\big(\tfrac{1}{z}\big)
+\int dz'^{-1}\ P\frac{i}{z'^{-1}}\ 
\Delta_G^\alpha
\big(\tfrac{1}{z},\tfrac{1}{z}{-}\tfrac{1}{z'}\big)\ ,\displaybreak[0]\\
\widetilde{\bar\Phi}{}_\partial^\alpha(x)
=\bar\Phi{}_D^\alpha(x)
-\int dx'\ P\frac{i}{x'}\ 
\bar\Phi{}_G^\alpha(x,x{-}x')\ ,\displaybreak[0]\\
\widetilde{\bar\Delta}{}_\partial^\alpha
\big(\tfrac{1}{z}\big)
=\bar\Delta{}_D^\alpha\big(\tfrac{1}{z}\big)
+\int dz'^{-1}\ P\frac{i}{z'^{-1}}\ 
\bar\Delta{}_G^\alpha
\big(\tfrac{1}{z},\tfrac{1}{z}{-}\tfrac{1}{z'}\big)\ ,\displaybreak[0]\\
\widetilde\Gamma{}_\partial^\alpha(x)
=\Gamma_D^\alpha(x)
-\int dx'\ P\frac{i}{x'}\,\Gamma_G^{(f)}{}^\alpha(x,x{-}x')\ ,\label{appelsap}\\
\widetilde{\widehat\Gamma}{}_\partial^\alpha\big(\tfrac{1}{z}\big)
=\widehat\Gamma{}_D^\alpha(x)
+\int dz'^{-1}\ P\frac{i}{z'^{-1}}\,\widehat\Gamma_G^{(f)}{}^\alpha
\big(\tfrac{1}{z},\tfrac{1}{z}{-}\tfrac{1}{z'}\big)\ .
\end{gather}
\end{subequations}

\bibliographystyle{apsrev}
\bibliography{references}

\begin{thebibliography}{51}
\expandafter\ifx\csname natexlab\endcsname\relax\def\natexlab#1{#1}\fi
\expandafter\ifx\csname bibnamefont\endcsname\relax
  \def\bibnamefont#1{#1}\fi
\expandafter\ifx\csname bibfnamefont\endcsname\relax
  \def\bibfnamefont#1{#1}\fi
\expandafter\ifx\csname citenamefont\endcsname\relax
  \def\citenamefont#1{#1}\fi
\expandafter\ifx\csname url\endcsname\relax
  \def\url#1{\texttt{#1}}\fi
\expandafter\ifx\csname urlprefix\endcsname\relax\def\urlprefix{URL }\fi
\providecommand{\bibinfo}[2]{#2}
\providecommand{\eprint}[2][]{\url{#2}}

\bibitem[{\citenamefont{Adams et~al.}(1991{\natexlab{a}})}]{Adams:1991rw}
\bibinfo{author}{\bibfnamefont{D.~L.} \bibnamefont{Adams}} \bibnamefont{et~al.}
  (\bibinfo{collaboration}{E581}), \bibinfo{journal}{Phys. Lett.}
  \textbf{\bibinfo{volume}{B261}}, \bibinfo{pages}{201}
  (\bibinfo{year}{1991}{\natexlab{a}}).

\bibitem[{\citenamefont{Adams et~al.}(1991{\natexlab{b}})}]{Adams:1991cs}
\bibinfo{author}{\bibfnamefont{D.~L.} \bibnamefont{Adams}} \bibnamefont{et~al.}
  (\bibinfo{collaboration}{FNAL-E704}), \bibinfo{journal}{Phys. Lett.}
  \textbf{\bibinfo{volume}{B264}}, \bibinfo{pages}{462}
  (\bibinfo{year}{1991}{\natexlab{b}}).

\bibitem[{\citenamefont{Bravar et~al.}(1996)}]{Bravar:1996ki}
\bibinfo{author}{\bibfnamefont{A.}~\bibnamefont{Bravar}} \bibnamefont{et~al.}
  (\bibinfo{collaboration}{Fermilab E704}), \bibinfo{journal}{Phys. Rev. Lett.}
  \textbf{\bibinfo{volume}{77}}, \bibinfo{pages}{2626} (\bibinfo{year}{1996}).

\bibitem[{\citenamefont{Airapetian et~al.}(2001)}]{Airapetian:2001eg}
\bibinfo{author}{\bibfnamefont{A.}~\bibnamefont{Airapetian}}
  \bibnamefont{et~al.} (\bibinfo{collaboration}{HERMES}),
  \bibinfo{journal}{Phys. Rev.} \textbf{\bibinfo{volume}{D64}},
  \bibinfo{pages}{097101} (\bibinfo{year}{2001}), \eprint{hep-ex/0104005}.

\bibitem[{\citenamefont{Adler et~al.}(2003)}]{Adler:2003pb}
\bibinfo{author}{\bibfnamefont{S.~S.} \bibnamefont{Adler}} \bibnamefont{et~al.}
  (\bibinfo{collaboration}{PHENIX}), \bibinfo{journal}{Phys. Rev. Lett.}
  \textbf{\bibinfo{volume}{91}}, \bibinfo{pages}{241803}
  (\bibinfo{year}{2003}), \eprint{hep-ex/0304038}.

\bibitem[{\citenamefont{Adams et~al.}(2004)}]{Adams:2003fx}
\bibinfo{author}{\bibfnamefont{J.}~\bibnamefont{Adams}} \bibnamefont{et~al.}
  (\bibinfo{collaboration}{STAR}), \bibinfo{journal}{Phys. Rev. Lett.}
  \textbf{\bibinfo{volume}{92}}, \bibinfo{pages}{171801}
  (\bibinfo{year}{2004}), \eprint{hep-ex/0310058}.

\bibitem[{\citenamefont{Airapetian et~al.}(2005)}]{Airapetian:2004tw}
\bibinfo{author}{\bibfnamefont{A.}~\bibnamefont{Airapetian}}
  \bibnamefont{et~al.} (\bibinfo{collaboration}{HERMES}),
  \bibinfo{journal}{Phys. Rev. Lett.} \textbf{\bibinfo{volume}{94}},
  \bibinfo{pages}{012002} (\bibinfo{year}{2005}), \eprint{hep-ex/0408013}.

\bibitem[{\citenamefont{Belitsky et~al.}(2003)\citenamefont{Belitsky, Ji, and
  Yuan}}]{Belitsky:2002sm}
\bibinfo{author}{\bibfnamefont{A.~V.} \bibnamefont{Belitsky}},
  \bibinfo{author}{\bibfnamefont{X.}~\bibnamefont{Ji}}, \bibnamefont{and}
  \bibinfo{author}{\bibfnamefont{F.}~\bibnamefont{Yuan}},
  \bibinfo{journal}{Nucl. Phys.} \textbf{\bibinfo{volume}{B656}},
  \bibinfo{pages}{165} (\bibinfo{year}{2003}), \eprint{hep-ph/0208038}.

\bibitem[{\citenamefont{Boer et~al.}(2003)\citenamefont{Boer, Mulders, and
  Pijlman}}]{Boer:2003cm}
\bibinfo{author}{\bibfnamefont{D.}~\bibnamefont{Boer}},
  \bibinfo{author}{\bibfnamefont{P.~J.} \bibnamefont{Mulders}},
  \bibnamefont{and} \bibinfo{author}{\bibfnamefont{F.}~\bibnamefont{Pijlman}},
  \bibinfo{journal}{Nucl. Phys.} \textbf{\bibinfo{volume}{B667}},
  \bibinfo{pages}{201} (\bibinfo{year}{2003}), \eprint{hep-ph/0303034}.

\bibitem[{\citenamefont{Collins}(1993)}]{Collins:1992kk}
\bibinfo{author}{\bibfnamefont{J.~C.} \bibnamefont{Collins}},
  \bibinfo{journal}{Nucl. Phys.} \textbf{\bibinfo{volume}{B396}},
  \bibinfo{pages}{161} (\bibinfo{year}{1993}), \eprint{hep-ph/9208213}.

\bibitem[{\citenamefont{Brodsky
  et~al.}(2002{\natexlab{a}})\citenamefont{Brodsky, Hwang, and
  Schmidt}}]{Brodsky:2002cx}
\bibinfo{author}{\bibfnamefont{S.~J.} \bibnamefont{Brodsky}},
  \bibinfo{author}{\bibfnamefont{D.~S.} \bibnamefont{Hwang}}, \bibnamefont{and}
  \bibinfo{author}{\bibfnamefont{I.}~\bibnamefont{Schmidt}},
  \bibinfo{journal}{Phys. Lett.} \textbf{\bibinfo{volume}{B530}},
  \bibinfo{pages}{99} (\bibinfo{year}{2002}{\natexlab{a}}),
  \eprint{hep-ph/0201296}.

\bibitem[{\citenamefont{Qiu and Sterman}(1999)}]{Qiu:1998ia}
\bibinfo{author}{\bibfnamefont{J.-w.} \bibnamefont{Qiu}} \bibnamefont{and}
  \bibinfo{author}{\bibfnamefont{G.}~\bibnamefont{Sterman}},
  \bibinfo{journal}{Phys. Rev.} \textbf{\bibinfo{volume}{D59}},
  \bibinfo{pages}{014004} (\bibinfo{year}{1999}), \eprint{hep-ph/9806356}.

\bibitem[{\citenamefont{Bomhof et~al.}(2004)\citenamefont{Bomhof, Mulders, and
  Pijlman}}]{Bomhof:2004aw}
\bibinfo{author}{\bibfnamefont{C.~J.} \bibnamefont{Bomhof}},
  \bibinfo{author}{\bibfnamefont{P.~J.} \bibnamefont{Mulders}},
  \bibnamefont{and} \bibinfo{author}{\bibfnamefont{F.}~\bibnamefont{Pijlman}},
  \bibinfo{journal}{Phys. Lett.} \textbf{\bibinfo{volume}{B596}},
  \bibinfo{pages}{277} (\bibinfo{year}{2004}), \eprint{hep-ph/0406099}.

\bibitem[{\citenamefont{Bacchetta et~al.}(2005)\citenamefont{Bacchetta, Bomhof,
  Mulders, and Pijlman}}]{Bacchetta:2005rm}
\bibinfo{author}{\bibfnamefont{A.}~\bibnamefont{Bacchetta}},
  \bibinfo{author}{\bibfnamefont{C.~J.} \bibnamefont{Bomhof}},
  \bibinfo{author}{\bibfnamefont{P.~J.} \bibnamefont{Mulders}},
  \bibnamefont{and} \bibinfo{author}{\bibfnamefont{F.}~\bibnamefont{Pijlman}},
  \bibinfo{journal}{Phys. Rev.} \textbf{\bibinfo{volume}{D72}},
  \bibinfo{pages}{034030} (\bibinfo{year}{2005}), \eprint{hep-ph/0505268}.

\bibitem[{\citenamefont{Bomhof et~al.}(2006)\citenamefont{Bomhof, Mulders, and
  Pijlman}}]{Bomhof:2006dp}
\bibinfo{author}{\bibfnamefont{C.~J.} \bibnamefont{Bomhof}},
  \bibinfo{author}{\bibfnamefont{P.~J.} \bibnamefont{Mulders}},
  \bibnamefont{and} \bibinfo{author}{\bibfnamefont{F.}~\bibnamefont{Pijlman}},
  \bibinfo{journal}{Eur. Phys. J.} \textbf{\bibinfo{volume}{C47}},
  \bibinfo{pages}{147} (\bibinfo{year}{2006}), \eprint{hep-ph/0601171}.

\bibitem[{\citenamefont{Collins}(2002)}]{Collins:2002kn}
\bibinfo{author}{\bibfnamefont{J.~C.} \bibnamefont{Collins}},
  \bibinfo{journal}{Phys. Lett.} \textbf{\bibinfo{volume}{B536}},
  \bibinfo{pages}{43} (\bibinfo{year}{2002}), \eprint{hep-ph/0204004}.

\bibitem[{\citenamefont{Brodsky
  et~al.}(2002{\natexlab{b}})\citenamefont{Brodsky, Hwang, and
  Schmidt}}]{Brodsky:2002rv}
\bibinfo{author}{\bibfnamefont{S.~J.} \bibnamefont{Brodsky}},
  \bibinfo{author}{\bibfnamefont{D.~S.} \bibnamefont{Hwang}}, \bibnamefont{and}
  \bibinfo{author}{\bibfnamefont{I.}~\bibnamefont{Schmidt}},
  \bibinfo{journal}{Nucl. Phys.} \textbf{\bibinfo{volume}{B642}},
  \bibinfo{pages}{344} (\bibinfo{year}{2002}{\natexlab{b}}),
  \eprint{hep-ph/0206259}.

\bibitem[{\citenamefont{Stratmann and Vogelsang}(1992)}]{Stratmann:1992gu}
\bibinfo{author}{\bibfnamefont{M.}~\bibnamefont{Stratmann}} \bibnamefont{and}
  \bibinfo{author}{\bibfnamefont{W.}~\bibnamefont{Vogelsang}},
  \bibinfo{journal}{Phys. Lett.} \textbf{\bibinfo{volume}{B295}},
  \bibinfo{pages}{277} (\bibinfo{year}{1992}).

\bibitem[{\citenamefont{Vogelsang}(1998)}]{Vogelsang:1998yd}
\bibinfo{author}{\bibfnamefont{W.}~\bibnamefont{Vogelsang}},
  \bibinfo{journal}{Acta Phys. Polon.} \textbf{\bibinfo{volume}{B29}},
  \bibinfo{pages}{1189} (\bibinfo{year}{1998}), \eprint{hep-ph/9805295}.

\bibitem[{\citenamefont{Collins and Metz}(2004)}]{Collins:2004nx}
\bibinfo{author}{\bibfnamefont{J.~C.} \bibnamefont{Collins}} \bibnamefont{and}
  \bibinfo{author}{\bibfnamefont{A.}~\bibnamefont{Metz}},
  \bibinfo{journal}{Phys. Rev. Lett.} \textbf{\bibinfo{volume}{93}},
  \bibinfo{pages}{252001} (\bibinfo{year}{2004}), \eprint{hep-ph/0408249}.

\bibitem[{\citenamefont{Mulders and Rodrigues}(2001)}]{Mulders:2000sh}
\bibinfo{author}{\bibfnamefont{P.~J.} \bibnamefont{Mulders}} \bibnamefont{and}
  \bibinfo{author}{\bibfnamefont{J.}~\bibnamefont{Rodrigues}},
  \bibinfo{journal}{Phys. Rev.} \textbf{\bibinfo{volume}{D63}},
  \bibinfo{pages}{094021} (\bibinfo{year}{2001}), \eprint{hep-ph/0009343}.

\bibitem[{\citenamefont{Vogelsang}()}]{Vogelsang}
\bibinfo{author}{\bibfnamefont{W.}~\bibnamefont{Vogelsang}},
  \bibinfo{note}{private communication}.

\bibitem[{\citenamefont{Boer and Vogelsang}(2004)}]{Boer:2003tx}
\bibinfo{author}{\bibfnamefont{D.}~\bibnamefont{Boer}} \bibnamefont{and}
  \bibinfo{author}{\bibfnamefont{W.}~\bibnamefont{Vogelsang}},
  \bibinfo{journal}{Phys. Rev.} \textbf{\bibinfo{volume}{D69}},
  \bibinfo{pages}{094025} (\bibinfo{year}{2004}), \eprint{hep-ph/0312320}.

\bibitem[{\citenamefont{Anselmino et~al.}(2006)\citenamefont{Anselmino,
  D'Alesio, Melis, and Murgia}}]{Anselmino:2006yq}
\bibinfo{author}{\bibfnamefont{M.}~\bibnamefont{Anselmino}},
  \bibinfo{author}{\bibfnamefont{U.}~\bibnamefont{D'Alesio}},
  \bibinfo{author}{\bibfnamefont{S.}~\bibnamefont{Melis}}, \bibnamefont{and}
  \bibinfo{author}{\bibfnamefont{F.}~\bibnamefont{Murgia}}
  (\bibinfo{year}{2006}), \eprint{hep-ph/0608211}.

\bibitem[{\citenamefont{Kouvaris et~al.}(2006)\citenamefont{Kouvaris, Qiu,
  Vogelsang, and Yuan}}]{Kouvaris:2006zy}
\bibinfo{author}{\bibfnamefont{C.}~\bibnamefont{Kouvaris}},
  \bibinfo{author}{\bibfnamefont{J.-W.} \bibnamefont{Qiu}},
  \bibinfo{author}{\bibfnamefont{W.}~\bibnamefont{Vogelsang}},
  \bibnamefont{and} \bibinfo{author}{\bibfnamefont{F.}~\bibnamefont{Yuan}},
  \bibinfo{journal}{Phys. Rev.} \textbf{\bibinfo{volume}{D74}},
  \bibinfo{pages}{114013} (\bibinfo{year}{2006}), \eprint{hep-ph/0609238}.

\bibitem[{\citenamefont{Gluck et~al.}(1998)\citenamefont{Gluck, Reya, and
  Vogt}}]{Gluck:1998xa}
\bibinfo{author}{\bibfnamefont{M.}~\bibnamefont{Gluck}},
  \bibinfo{author}{\bibfnamefont{E.}~\bibnamefont{Reya}}, \bibnamefont{and}
  \bibinfo{author}{\bibfnamefont{A.}~\bibnamefont{Vogt}},
  \bibinfo{journal}{Eur. Phys. J.} \textbf{\bibinfo{volume}{C5}},
  \bibinfo{pages}{461} (\bibinfo{year}{1998}), \eprint{hep-ph/9806404}.

\bibitem[{\citenamefont{Leader et~al.}(2006)\citenamefont{Leader, Sidorov, and
  Stamenov}}]{Leader:2005ci}
\bibinfo{author}{\bibfnamefont{E.}~\bibnamefont{Leader}},
  \bibinfo{author}{\bibfnamefont{A.~V.} \bibnamefont{Sidorov}},
  \bibnamefont{and} \bibinfo{author}{\bibfnamefont{D.~B.}
  \bibnamefont{Stamenov}}, \bibinfo{journal}{Phys. Rev.}
  \textbf{\bibinfo{volume}{D73}}, \bibinfo{pages}{034023}
  (\bibinfo{year}{2006}), \eprint{hep-ph/0512114}.

\bibitem[{\citenamefont{Hirai et~al.}(2006)\citenamefont{Hirai, Kumano, and
  Saito}}]{Hirai:2006sr}
\bibinfo{author}{\bibfnamefont{M.}~\bibnamefont{Hirai}},
  \bibinfo{author}{\bibfnamefont{S.}~\bibnamefont{Kumano}}, \bibnamefont{and}
  \bibinfo{author}{\bibfnamefont{N.}~\bibnamefont{Saito}},
  \bibinfo{journal}{Phys. Rev.} \textbf{\bibinfo{volume}{D74}},
  \bibinfo{pages}{014015} (\bibinfo{year}{2006}), \eprint{hep-ph/0603213}.

\bibitem[{\citenamefont{Thorne et~al.}(2006)\citenamefont{Thorne, Martin, and
  Stirling}}]{Thorne:2006zu}
\bibinfo{author}{\bibfnamefont{R.~S.} \bibnamefont{Thorne}},
  \bibinfo{author}{\bibfnamefont{A.~D.} \bibnamefont{Martin}},
  \bibnamefont{and} \bibinfo{author}{\bibfnamefont{W.~J.}
  \bibnamefont{Stirling}} (\bibinfo{year}{2006}), \eprint{hep-ph/0606244}.

\bibitem[{\citenamefont{Signal and Thomas}(1989)}]{Signal:1989yc}
\bibinfo{author}{\bibfnamefont{A.~I.} \bibnamefont{Signal}} \bibnamefont{and}
  \bibinfo{author}{\bibfnamefont{A.~W.} \bibnamefont{Thomas}},
  \bibinfo{journal}{Phys. Rev.} \textbf{\bibinfo{volume}{D40}},
  \bibinfo{pages}{2832} (\bibinfo{year}{1989}).

\bibitem[{\citenamefont{Barone and Drago}(1993)}]{Barone:1993iq}
\bibinfo{author}{\bibfnamefont{V.}~\bibnamefont{Barone}} \bibnamefont{and}
  \bibinfo{author}{\bibfnamefont{A.}~\bibnamefont{Drago}},
  \bibinfo{journal}{Nucl. Phys.} \textbf{\bibinfo{volume}{A552}},
  \bibinfo{pages}{479} (\bibinfo{year}{1993}).

\bibitem[{\citenamefont{Jakob et~al.}(1997)\citenamefont{Jakob, Mulders, and
  Rodrigues}}]{Jakob:1997wg}
\bibinfo{author}{\bibfnamefont{R.}~\bibnamefont{Jakob}},
  \bibinfo{author}{\bibfnamefont{P.~J.} \bibnamefont{Mulders}},
  \bibnamefont{and}
  \bibinfo{author}{\bibfnamefont{J.}~\bibnamefont{Rodrigues}},
  \bibinfo{journal}{Nucl. Phys.} \textbf{\bibinfo{volume}{A626}},
  \bibinfo{pages}{937} (\bibinfo{year}{1997}), \eprint{hep-ph/9704335}.

\bibitem[{\citenamefont{Diakonov et~al.}(1998)\citenamefont{Diakonov, Petrov,
  Pobylitsa, Polyakov, and Weiss}}]{Diakonov:1998ze}
\bibinfo{author}{\bibfnamefont{D.}~\bibnamefont{Diakonov}},
  \bibinfo{author}{\bibfnamefont{V.~Y.} \bibnamefont{Petrov}},
  \bibinfo{author}{\bibfnamefont{P.~V.} \bibnamefont{Pobylitsa}},
  \bibinfo{author}{\bibfnamefont{M.~V.} \bibnamefont{Polyakov}},
  \bibnamefont{and} \bibinfo{author}{\bibfnamefont{C.}~\bibnamefont{Weiss}},
  \bibinfo{journal}{Phys. Rev.} \textbf{\bibinfo{volume}{D58}},
  \bibinfo{pages}{038502} (\bibinfo{year}{1998}).

\bibitem[{\citenamefont{Goeke et~al.}(2001)\citenamefont{Goeke, Pobylitsa,
  Polyakov, Schweitzer, and Urbano}}]{Goeke:2000wv}
\bibinfo{author}{\bibfnamefont{K.}~\bibnamefont{Goeke}},
  \bibinfo{author}{\bibfnamefont{P.~V.} \bibnamefont{Pobylitsa}},
  \bibinfo{author}{\bibfnamefont{M.~V.} \bibnamefont{Polyakov}},
  \bibinfo{author}{\bibfnamefont{P.}~\bibnamefont{Schweitzer}},
  \bibnamefont{and} \bibinfo{author}{\bibfnamefont{D.}~\bibnamefont{Urbano}},
  \bibinfo{journal}{Acta Phys. Polon.} \textbf{\bibinfo{volume}{B32}},
  \bibinfo{pages}{1201} (\bibinfo{year}{2001}), \eprint{hep-ph/0001272}.

\bibitem[{\citenamefont{Gluck et~al.}(2001)\citenamefont{Gluck, Reya,
  Stratmann, and Vogelsang}}]{Gluck:2000dy}
\bibinfo{author}{\bibfnamefont{M.}~\bibnamefont{Gluck}},
  \bibinfo{author}{\bibfnamefont{E.}~\bibnamefont{Reya}},
  \bibinfo{author}{\bibfnamefont{M.}~\bibnamefont{Stratmann}},
  \bibnamefont{and}
  \bibinfo{author}{\bibfnamefont{W.}~\bibnamefont{Vogelsang}},
  \bibinfo{journal}{Phys. Rev.} \textbf{\bibinfo{volume}{D63}},
  \bibinfo{pages}{094005} (\bibinfo{year}{2001}), \eprint{hep-ph/0011215}.

\bibitem[{\citenamefont{Bacchetta et~al.}(2001)\citenamefont{Bacchetta, Kundu,
  Metz, and Mulders}}]{Bacchetta:2001di}
\bibinfo{author}{\bibfnamefont{A.}~\bibnamefont{Bacchetta}},
  \bibinfo{author}{\bibfnamefont{R.}~\bibnamefont{Kundu}},
  \bibinfo{author}{\bibfnamefont{A.}~\bibnamefont{Metz}}, \bibnamefont{and}
  \bibinfo{author}{\bibfnamefont{P.}~\bibnamefont{Mulders}},
  \bibinfo{journal}{Phys. Lett.} \textbf{\bibinfo{volume}{B506}},
  \bibinfo{pages}{155} (\bibinfo{year}{2001}), \eprint{hep-ph/0102278}.

\bibitem[{\citenamefont{Wakamatsu}(2001)}]{Wakamatsu:2000fd}
\bibinfo{author}{\bibfnamefont{M.}~\bibnamefont{Wakamatsu}},
  \bibinfo{journal}{Phys. Lett.} \textbf{\bibinfo{volume}{B509}},
  \bibinfo{pages}{59} (\bibinfo{year}{2001}), \eprint{hep-ph/0012331}.

\bibitem[{\citenamefont{Bacchetta et~al.}(2002)\citenamefont{Bacchetta, Kundu,
  Metz, and Mulders}}]{Bacchetta:2002tk}
\bibinfo{author}{\bibfnamefont{A.}~\bibnamefont{Bacchetta}},
  \bibinfo{author}{\bibfnamefont{R.}~\bibnamefont{Kundu}},
  \bibinfo{author}{\bibfnamefont{A.}~\bibnamefont{Metz}}, \bibnamefont{and}
  \bibinfo{author}{\bibfnamefont{P.}~\bibnamefont{Mulders}},
  \bibinfo{journal}{Phys. Rev.} \textbf{\bibinfo{volume}{D65}},
  \bibinfo{pages}{094021} (\bibinfo{year}{2002}), \eprint{hep-ph/0201091}.

\bibitem[{\citenamefont{Yang}(2002)}]{Yang:2002gh}
\bibinfo{author}{\bibfnamefont{J.-J.} \bibnamefont{Yang}},
  \bibinfo{journal}{Phys. Rev.} \textbf{\bibinfo{volume}{D65}},
  \bibinfo{pages}{094035} (\bibinfo{year}{2002}).

\bibitem[{\citenamefont{Efremov}(2004)}]{Efremov:2004ph}
\bibinfo{author}{\bibfnamefont{A.~V.} \bibnamefont{Efremov}},
  \bibinfo{journal}{Annalen Phys.} \textbf{\bibinfo{volume}{13}},
  \bibinfo{pages}{651} (\bibinfo{year}{2004}), \eprint{hep-ph/0410389}.

\bibitem[{\citenamefont{Anselmino et~al.}(1995)\citenamefont{Anselmino,
  Boglione, and Murgia}}]{Anselmino:1994tv}
\bibinfo{author}{\bibfnamefont{M.}~\bibnamefont{Anselmino}},
  \bibinfo{author}{\bibfnamefont{M.}~\bibnamefont{Boglione}}, \bibnamefont{and}
  \bibinfo{author}{\bibfnamefont{F.}~\bibnamefont{Murgia}},
  \bibinfo{journal}{Phys. Lett.} \textbf{\bibinfo{volume}{B362}},
  \bibinfo{pages}{164} (\bibinfo{year}{1995}), \eprint{hep-ph/9503290}.

\bibitem[{\citenamefont{Anselmino and Murgia}(1998)}]{Anselmino:1998yz}
\bibinfo{author}{\bibfnamefont{M.}~\bibnamefont{Anselmino}} \bibnamefont{and}
  \bibinfo{author}{\bibfnamefont{F.}~\bibnamefont{Murgia}},
  \bibinfo{journal}{Phys. Lett.} \textbf{\bibinfo{volume}{B442}},
  \bibinfo{pages}{470} (\bibinfo{year}{1998}), \eprint{hep-ph/9808426}.

\bibitem[{\citenamefont{Soper}(1977)}]{Soper:1976jc}
\bibinfo{author}{\bibfnamefont{D.~E.} \bibnamefont{Soper}},
  \bibinfo{journal}{Phys. Rev.} \textbf{\bibinfo{volume}{D15}},
  \bibinfo{pages}{1141} (\bibinfo{year}{1977}).

\bibitem[{\citenamefont{Soper}(1979)}]{Soper:1979fq}
\bibinfo{author}{\bibfnamefont{D.~E.} \bibnamefont{Soper}},
  \bibinfo{journal}{Phys. Rev. Lett.} \textbf{\bibinfo{volume}{43}},
  \bibinfo{pages}{1847} (\bibinfo{year}{1979}).

\bibitem[{\citenamefont{Ralston and Soper}(1979)}]{Ralston:1979ys}
\bibinfo{author}{\bibfnamefont{J.~P.} \bibnamefont{Ralston}} \bibnamefont{and}
  \bibinfo{author}{\bibfnamefont{D.~E.} \bibnamefont{Soper}},
  \bibinfo{journal}{Nucl. Phys.} \textbf{\bibinfo{volume}{B152}},
  \bibinfo{pages}{109} (\bibinfo{year}{1979}).

\bibitem[{\citenamefont{Collins and Soper}(1982)}]{Collins:1981uw}
\bibinfo{author}{\bibfnamefont{J.~C.} \bibnamefont{Collins}} \bibnamefont{and}
  \bibinfo{author}{\bibfnamefont{D.~E.} \bibnamefont{Soper}},
  \bibinfo{journal}{Nucl. Phys.} \textbf{\bibinfo{volume}{B194}},
  \bibinfo{pages}{445} (\bibinfo{year}{1982}).

\bibitem[{\citenamefont{Collins et~al.}(1982)\citenamefont{Collins, Soper, and
  Sterman}}]{Collins:1981tt}
\bibinfo{author}{\bibfnamefont{J.~C.} \bibnamefont{Collins}},
  \bibinfo{author}{\bibfnamefont{D.~E.} \bibnamefont{Soper}}, \bibnamefont{and}
  \bibinfo{author}{\bibfnamefont{G.}~\bibnamefont{Sterman}},
  \bibinfo{journal}{Phys. Lett.} \textbf{\bibinfo{volume}{B109}},
  \bibinfo{pages}{388} (\bibinfo{year}{1982}).

\bibitem[{\citenamefont{Collins et~al.}(1983)\citenamefont{Collins, Soper, and
  Sterman}}]{Collins:1982wa}
\bibinfo{author}{\bibfnamefont{J.~C.} \bibnamefont{Collins}},
  \bibinfo{author}{\bibfnamefont{D.~E.} \bibnamefont{Soper}}, \bibnamefont{and}
  \bibinfo{author}{\bibfnamefont{G.}~\bibnamefont{Sterman}},
  \bibinfo{journal}{Nucl. Phys.} \textbf{\bibinfo{volume}{B223}},
  \bibinfo{pages}{381} (\bibinfo{year}{1983}).

\bibitem[{\citenamefont{Jaffe and Ji}(1992)}]{Jaffe:1991ra}
\bibinfo{author}{\bibfnamefont{R.~L.} \bibnamefont{Jaffe}} \bibnamefont{and}
  \bibinfo{author}{\bibfnamefont{X.-D.} \bibnamefont{Ji}},
  \bibinfo{journal}{Nucl. Phys.} \textbf{\bibinfo{volume}{B375}},
  \bibinfo{pages}{527} (\bibinfo{year}{1992}).

\bibitem[{\citenamefont{Mulders and Tangerman}(1996)}]{Mulders:1995dh}
\bibinfo{author}{\bibfnamefont{P.~J.} \bibnamefont{Mulders}} \bibnamefont{and}
  \bibinfo{author}{\bibfnamefont{R.~D.} \bibnamefont{Tangerman}},
  \bibinfo{journal}{Nucl. Phys.} \textbf{\bibinfo{volume}{B461}},
  \bibinfo{pages}{197} (\bibinfo{year}{1996}), \eprint{hep-ph/9510301}.

\bibitem[{\citenamefont{Bacchetta and Radici}(2004)}]{Bacchetta:2004it}
\bibinfo{author}{\bibfnamefont{A.}~\bibnamefont{Bacchetta}} \bibnamefont{and}
  \bibinfo{author}{\bibfnamefont{M.}~\bibnamefont{Radici}},
  \bibinfo{journal}{Phys. Rev.} \textbf{\bibinfo{volume}{D70}},
  \bibinfo{pages}{094032} (\bibinfo{year}{2004}), \eprint{hep-ph/0409174}.

\end{thebibliography}

\clearpage

\begin{table}
\centering
\begin{minipage}[t]{0.45\textwidth}
\centering
\includegraphics{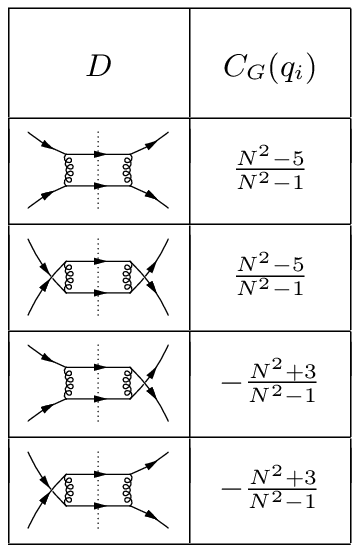}
\caption{Diagrams that contribute to $qq{\rightarrow}qq$ scattering 
(first column) 
and gluonic pole strengths of the incoming quarks (second column),
where $N$ is the number of colors.
For each diagram the gluonic pole strengths of the outgoing quarks are given by
$C_G(q_f){=}{-}C_G(q_i)$.\label{Tqq2qq}}
\end{minipage}\hspace{5mm}
\begin{minipage}[t]{0.45\textwidth}
\centering
\includegraphics{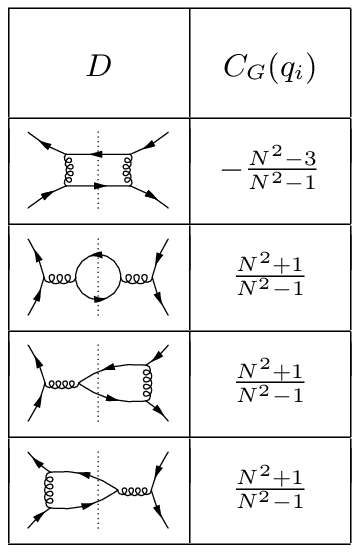}
\caption{Diagrams that contribute to $q\bar q{\rightarrow}q\bar q$ scattering 
(first column)
and gluonic pole strengths of the incoming quarks (second column).
The gluonic pole strengths of the other (anti)quarks are given by
$C_G(q_f){=}C_G(\bar q_f){=}{-}C_G(\bar q_i){=}{-}C_G(q_i)$.
\label{Tqq_2qq_}}
\end{minipage}\\[1cm]
\begin{minipage}[t]{0.45\textwidth}
\centering
\includegraphics{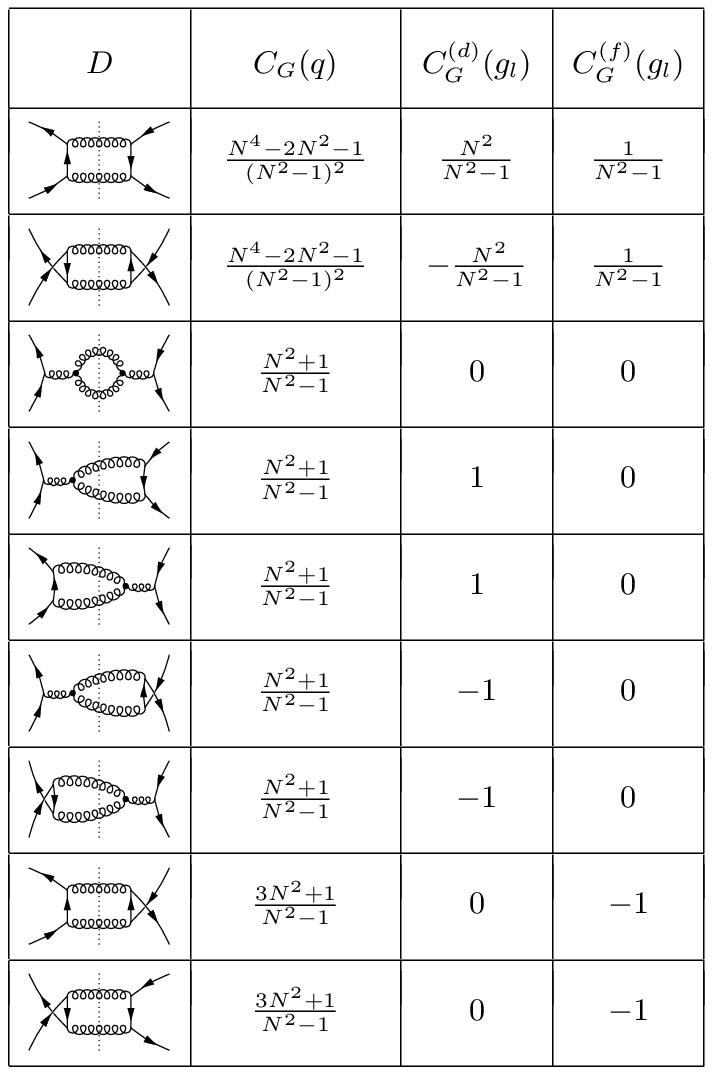}
\caption{Diagrams that contribute to $q\bar q{\rightarrow}gg$ scattering 
(first column) and the gluonic pole strengths of the incoming quarks (second column)
and the lower glouns ($g_l$) in the final state (third and fourth column).
The gluonic pole strengths of the incoming antiquarks and the upper gluons ($g_u$) in the final state are connected to those given above through the relations 
$C_G(\bar q){=}C_G(q)$,
$C_G^{(d)}(g_u){=}{-}C_G^{(d)}(g_l)$ and 
$C_G^{(f)}(g_u){=}C_G^{(f)}(g_l)$.\label{Tqq_2gg}}
\end{minipage}\hspace{5mm}
\begin{minipage}[t]{0.45\textwidth}
\centering
\includegraphics{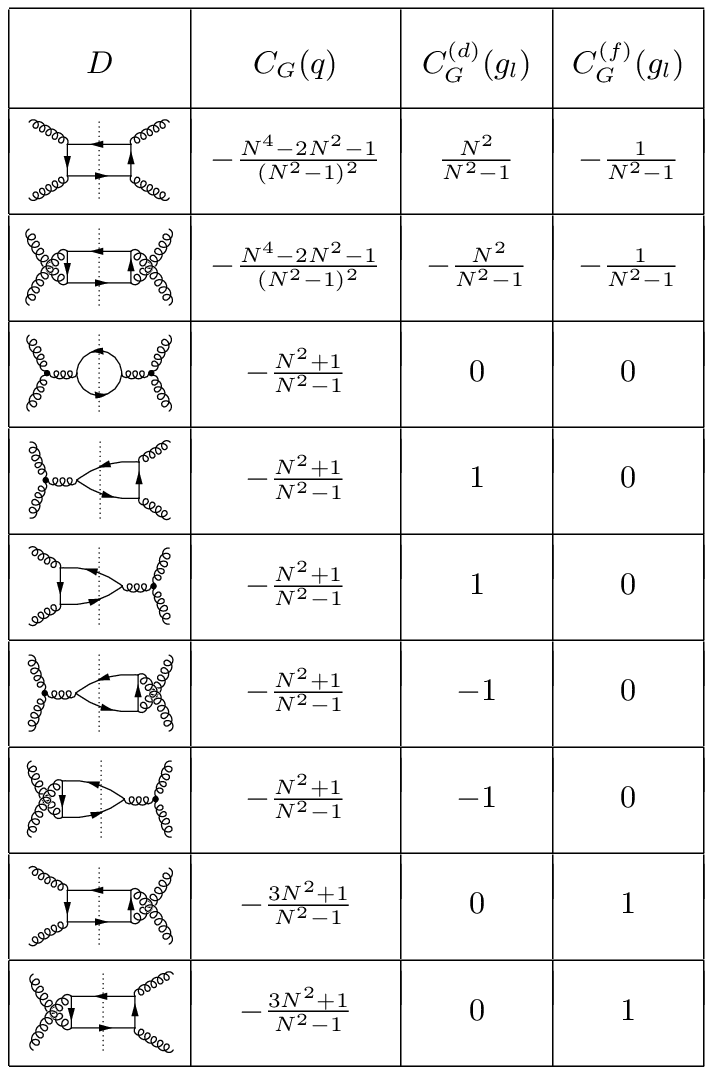}
\caption{Diagrams that contribute to $gg{\rightarrow}q\bar q$ scattering 
(first column) and the gluonic pole strengths of the outgoing quarks (second column)
and the lower glouns ($g_l$) in the initial state (third and fourth column).
The gluonic pole strengths of the outgoing antiquarks and the upper gluons ($g_u$) in the initial state are connected to those given above through the relations 
$C_G(\bar q){=}C_G(q)$,
$C_G^{(d)}(g_u){=}{-}C_G^{(d)}(g_l)$ and 
$C_G^{(f)}(g_u){=}C_G^{(f)}(g_l)$.\label{Tgg2qq_}}
\end{minipage}
\end{table}

\begin{table}
\centering
\begin{minipage}{\textwidth}
\centering
\begin{minipage}[t]{0.46\textwidth}
\centering
\includegraphics{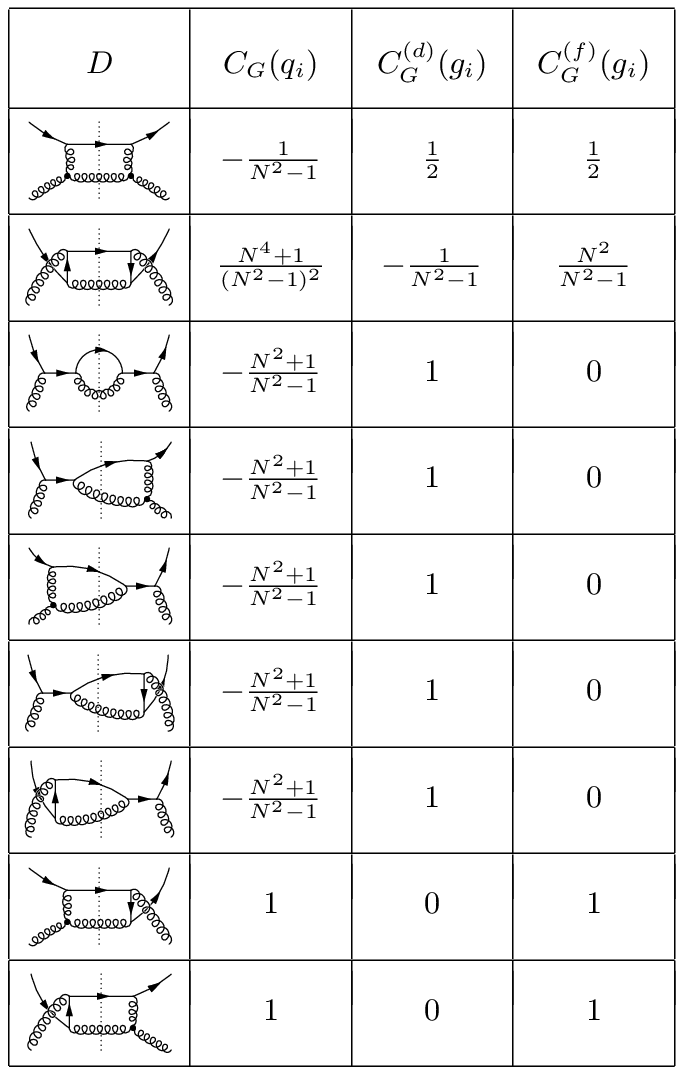}
\caption{Diagrams that contribute to $qg{\rightarrow}qg$ scattering 
(first column) and the gluonic pole strengths of the incoming quarks (second column) 
and gluons (third and fourth column).
The gluonic pole strengths of the outgoing partons are connected to those given above through the relations
$C_G(q_f){=}{-}C_G(q_i)$, $C_G^{(d)}(g_f){=}C_G^{(d)}(g_i)$ and $C_G^{(f)}(g_f){=}{-}C_G^{(f)}(g_i)$.\label{Tqg2qg}}
\end{minipage}\hspace{\stretch{1}}
\begin{minipage}[t]{0.46\textwidth}
\centering
\includegraphics{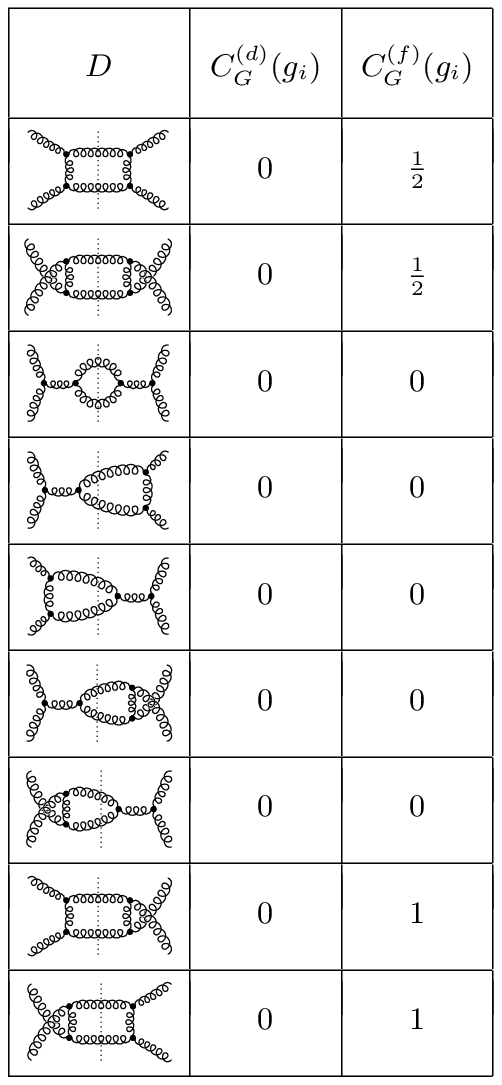}
\caption{The diagrams without 4-gluon vertices that contribute to $gg{\rightarrow}gg$ scattering 
(first column) and the gluonic pole strengths of the incoming glouns 
(second and third column).
The gluonic pole strengths of the outgoing gluons are connected to those given above through the relations 
$C_G^{(d)}(g_f){=}C_G^{(d)}(g_i)$ and 
$C_G^{(f)}(g_f){=}{-}C_G^{(f)}(g_i)$.\label{Tgg2gg}}
\end{minipage}\\[7mm]
\begin{minipage}[t]{0.47\textwidth}
\centering
\includegraphics{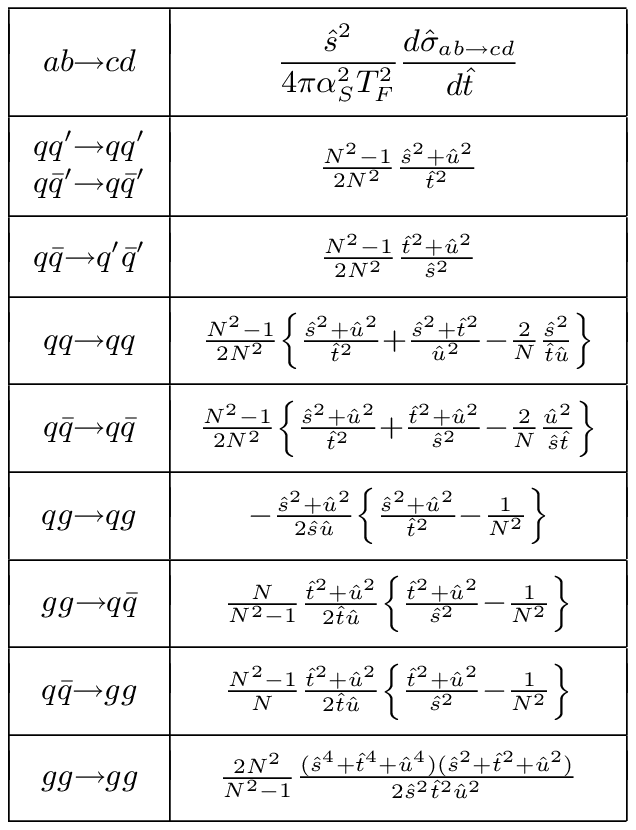}
\caption{Unpolarized cross sections as defined in section~\ref{MainText}.\label{PCSunpolarized}}
\end{minipage}\hspace{\stretch{1}}
\begin{minipage}[t]{0.47\textwidth}
\centering
\includegraphics{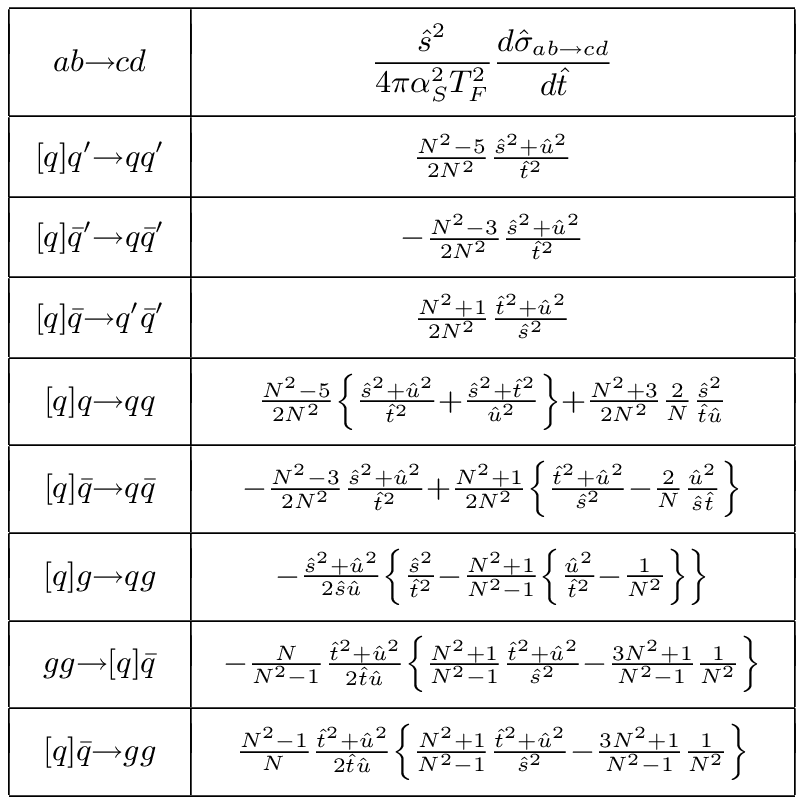}
\caption{Unpolarized gluonic pole cross sections as defined in section~\ref{MainText} when the gluonic pole is contributed by a quark.\label{GPCSunpolarized}}
\end{minipage}
\end{minipage}
\end{table}

\begin{table}
\centering
\begin{minipage}[t]{0.46\textwidth}
\centering
\includegraphics{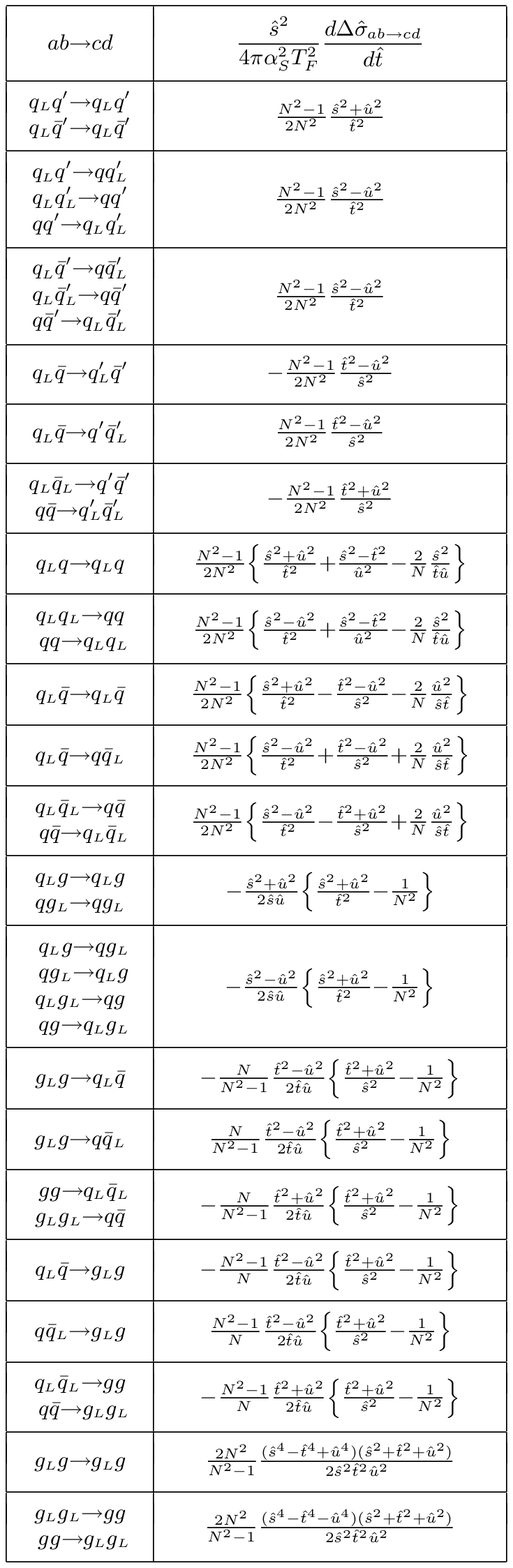}
\caption{Longitudinally polarized cross sections as defined in section~\ref{MainText}.
This reproduces the expressions in~\cite{Stratmann:1992gu} and adds the processes with both polarized partons in the initial or final state.\label{PCSlongitudinal}}
\end{minipage}\hspace{\stretch{1}}
\begin{minipage}[t]{0.48\textwidth}
\centering
\includegraphics{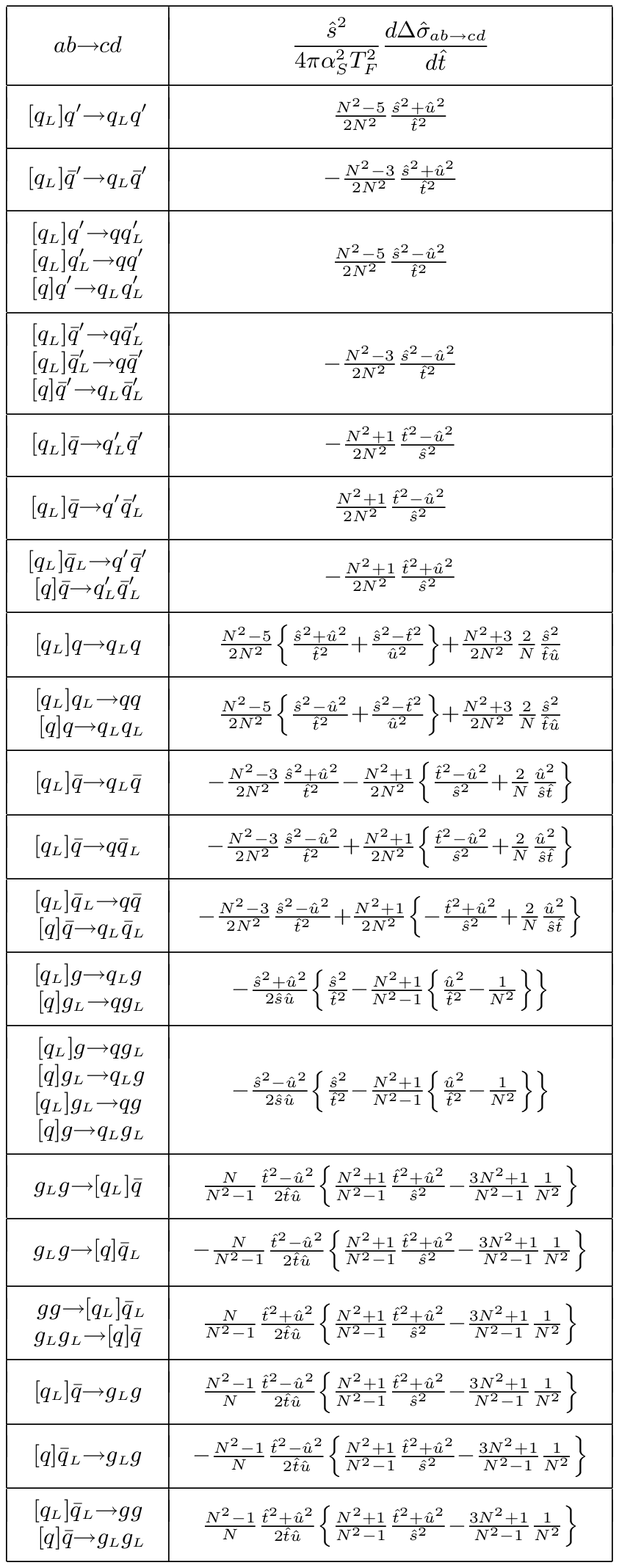}
\caption{Longitudinally polarized gluonic pole cross sections as defined in section~\ref{MainText} when the gluonic pole is contributed by a quark.\label{GPCSlongitudinal}}
\end{minipage}
\end{table}

\begin{table}
\centering
\begin{minipage}[t]{0.45\textwidth}
\centering
\includegraphics{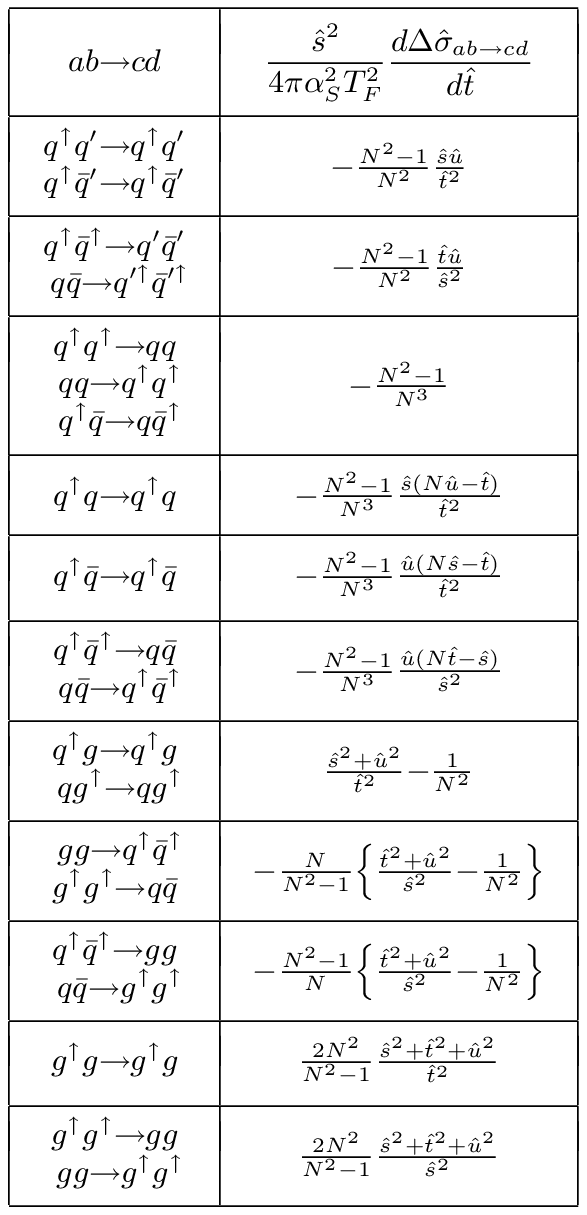}
\caption{Transversally polarized cross sections as defined in section~\ref{MainText}.
These reproduce the expressions in~\cite{Bacchetta:2004it},
except for an overall sign difference in the processes 
$q^\uparrow g{\rightarrow}q^\uparrow g$ and 
$qg^\uparrow{\rightarrow}qg^\uparrow$.\label{PCStransverse}}
\end{minipage}\hspace{1cm}
\begin{minipage}[t]{0.45\textwidth}
\centering
\includegraphics{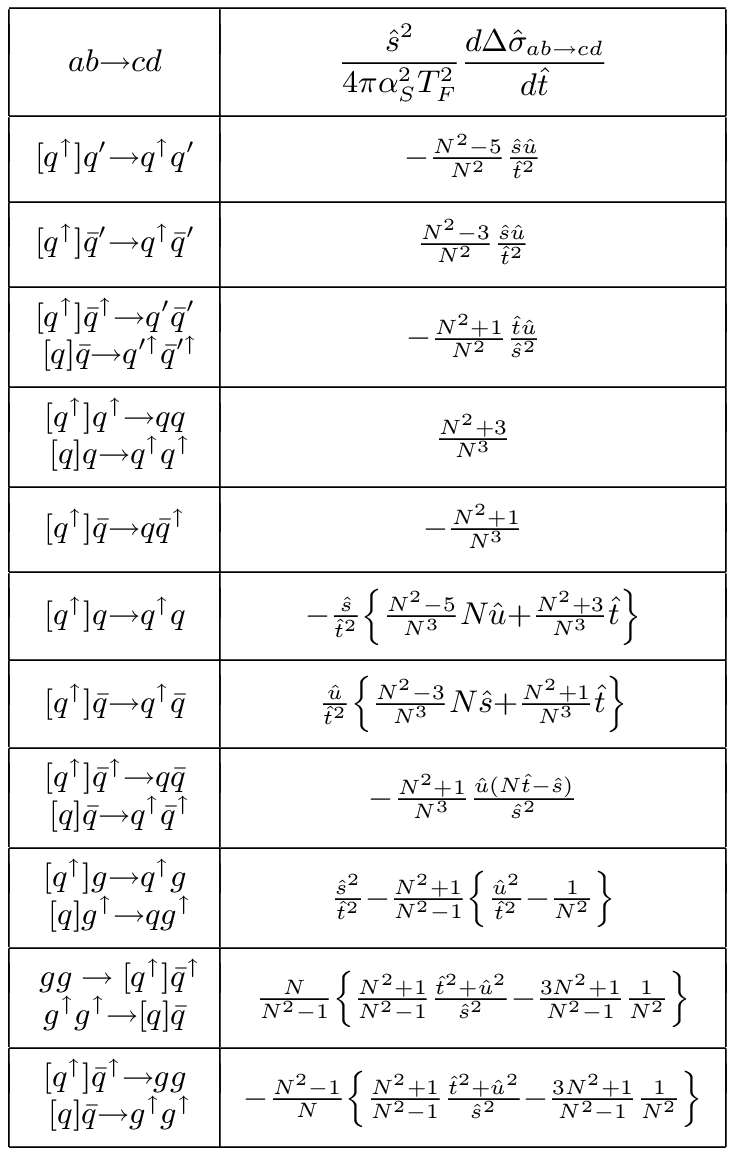}
\caption{Transversally polarized gluonic pole cross sections as defined in section~\ref{MainText} when the gluonic pole is contributed by a quark.\label{GPCStransverse}}
\end{minipage}\\[2cm]
\begin{minipage}{\textwidth}
\centering
\includegraphics{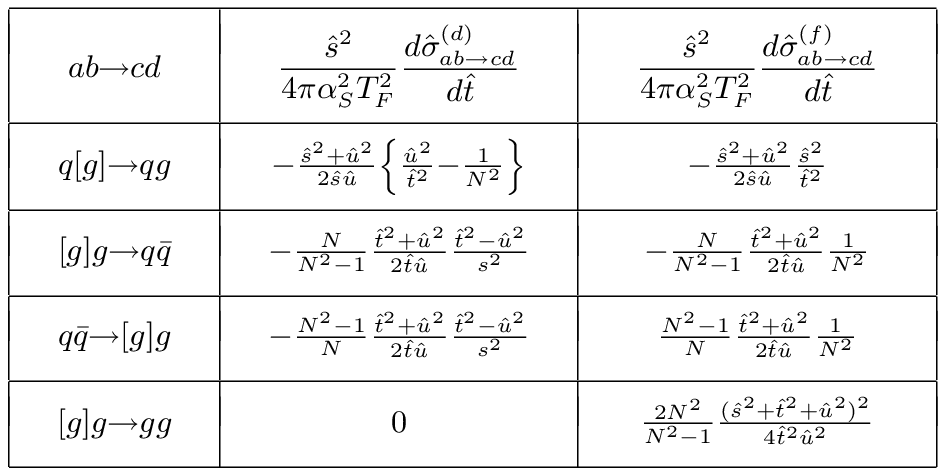}\\
\parbox{0.5\textwidth}{
\caption{Unpolarized gluonic pole cross sections as defined in section~\ref{MainText} when the gluonic pole is contributed by a gluon.\label{gGPCSunpolarized}}}
\end{minipage}
\end{table}

\begin{table}
\begin{minipage}[b]{0.9\textwidth}
\includegraphics{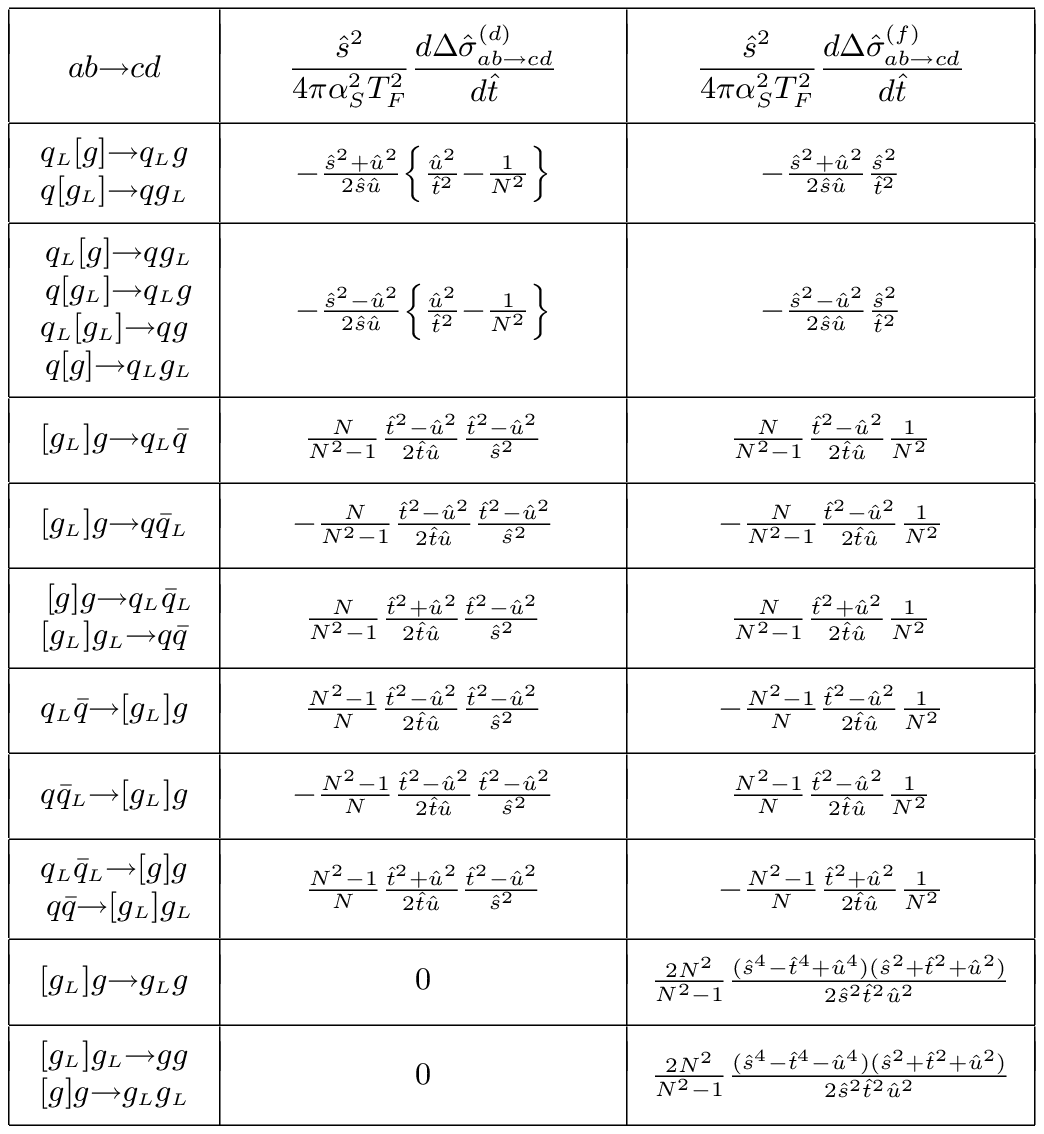}
\parbox{0.6\textwidth}{
\caption{Longitudinally polarized gluonic pole cross sections as defined in section~\ref{MainText} when the gluonic pole is contributed by a gluon.\label{gGPCSlongitudinal}}}\\
\vspace*{2cm}
\includegraphics{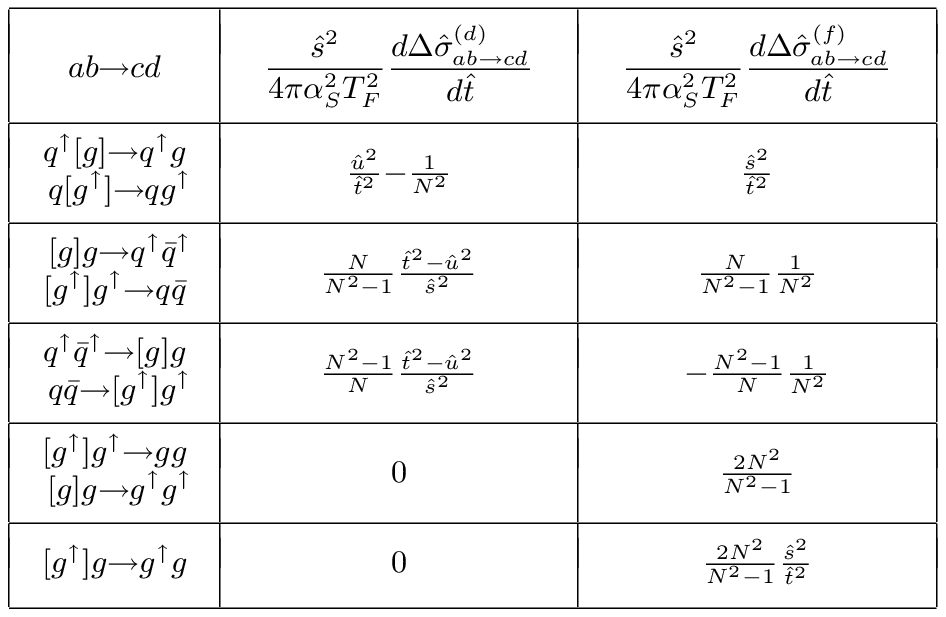}
\parbox{0.6\textwidth}{
\caption{Transversally polarized gluonic pole cross sections as defined in section~\ref{MainText} when the gluonic pole is contributed by a gluon.\label{gGPCStransverse}}}
\end{minipage}
\rule[-1cm]{0pt}{\textheight}
\end{table}

\end{document}